\documentclass[aps,nofootinbib,showpacs,preprintnumbers,amsmath,amssymb]{revtex4}
\usepackage{graphicx}% Include figure files
\usepackage{dcolumn}% Align table columns on decimal point
\usepackage{bm}% bold math
\usepackage{epsfig}
\usepackage{epsf}
\usepackage{amssymb}
\usepackage{amsmath}
\def\be{\begin{equation}}
\def\ee{\end{equation}}
\def\bea{\begin{eqnarray}}
\def\eea{\end{eqnarray}}
\def\bear{\begin{array}}
\def\eear{\end{array}}
\def\bes{\begin{subequations}}
\def\ees{\end{subequations}}
\newcommand{\MSbar}{\overline{\rm MS}}  

\newcommand{\A}{{\mathcal{A}}}
\newcommand{\tA}{{\widetilde {\mathcal{A}}}}
\newcommand{\ta}{{\widetilde a}}
\newcommand{\td}{{\widetilde d}}

\newcommand{\tk}{{\widetilde k}}

\newcommand{\ntH}{{\mathfrak A}}

\newcommand{\tal}{{\widetilde \alpha}}
\newcommand{\tu}{{\widetilde u}}
\usepackage{graphics}
\usepackage{graphicx}% Include figure files
\usepackage{dcolumn}% Align table columns on decimal point
\usepackage{bm}% bold math
\usepackage{epsfig}
\usepackage{graphicx}
\usepackage{multirow}
\usepackage{dcolumn}% Align table columns on decimal point
\usepackage{graphicx,epsfig}% 

\begin{document}
\preprint{USM-TH-315}

\title{Techniques of evaluation of QCD low-energy physical quantities with running coupling with infrared fixed point}

\author{Gorazd Cveti\v{c}}
%  \email{gorazd.cvetic@usm.cl}

\affiliation{Dept.~of Physics, Universidad T\'ecnica
Federico Santa Mar\'{\i}a (UTFSM), Casilla 110-V, Valpara\'{\i}so, Chile}

\date{\today}

\begin{abstract}
Perturbative QCD (pQCD) running coupling $a(Q^2)$ ($\equiv \alpha_s(Q^2)/\pi$)
is expected to get modified at low spacelike momenta 
$0 < Q^2 \alt 1 \ {\rm GeV}^2$ so that, 
instead of having unphysical (Landau) singularities 
it remains smooth and finite there, due to infared (IR) fixed point.
This behavior is suggested by: Gribov-Zwanziger approach,
Dyson-Schwinger equations (DSE) and other functional methods, 
lattice calculations, light-front holographic mapping
AdS/CFT modified by a dilaton background, and by
most of the analytic (holomorphic) QCD models. 
All such couplings, $\A(Q^2)$, differ from the pQCD
couplings $a(Q^2)$ at $|Q| \agt 1$ GeV 
by nonperturbative (NP) terms, typically by
some power-suppressed terms $\sim 1/Q^{2 N}$.

Evaluations of low-energy physical QCD quantities in terms of such $\A(Q^2)$
couplings (with IR fixed point) at a level beyond one-loop are usually performed
with (truncated) power series in $\A(Q^2)$. We argue that such an evaluation is not correct, because the NP terms in general get out of
control as the number of terms in the power series increases.
The series consequently become increasingly unstable under the variation
of the renormalization scale, and have a fast asymptotic
divergent behavior compounded by the renormalon problem. We argue that an alternative series
in terms of logarithmic derivatives of $\A(Q^2)$ should be used.
Further, a Pad\'e-related resummation based on this series gives results which
are renormalization scale independent and show very good convergence.
Timelike low-energy observables can be evaluated
analogously, using the integral transformation
which relates the timelike observable with the corresponding
spacelike observable.
\end{abstract}

\maketitle

\section{Introduction}
\label{sec:intro}

\begin{figure}[htb]
\includegraphics[width=55mm]{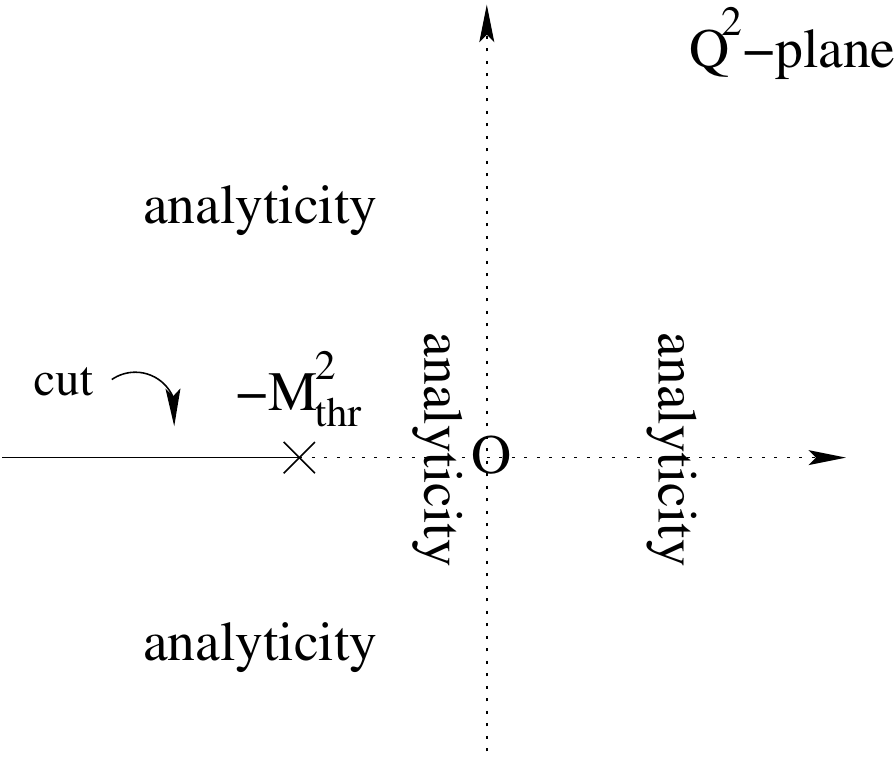}
\caption{{\footnotesize The typical region of analyticity of
a spacelike observable $d(Q^2)$ in the complex $Q^2$-plane.}}
\label{cut}
\end{figure}

One of the main problems in QCD is to understand the theory
at low (hadronic) scales $|q| \alt 1$ GeV. The usual perturbative
QCD (pQCD) coupling $a(Q^2)$ ($\equiv \alpha_s^{\rm (pt)}(Q^2)/\pi$),
where $q^2 \equiv -Q^2$ is the squared momentum transfer,
suffers from Landau singularities at low
scales: $|Q^2| \alt 1 \ {\rm GeV}^2$ and $Q^2 \not< 0$. 
These singularities can be called unphysical for the following
reason: the spacelike observables $d(Q^2)$, which are expected to
be evaluated as a function of $a(\kappa Q^2)$ (with $\kappa \sim 1$),
do not have such singularities. In fact, $d(Q^2)$ are analytic functions
of $Q^2$ in the entire complex plane with the exception of the
negative axis $Q^2 < - M_{\rm thr}^2$ (where $M_{\rm thr}^2 \sim 10^{-1} \ {\rm GeV}^2$
is a squared threshold scale), see Fig.~\ref{cut},
this property following from the basic
principles of quantum field theories such as locality,
unitarity and microcausality.

The mentioned analyticity properties of a realistic QCD coupling
$\A(Q^2)$ are supported by Gribov-Zwanziger approach
\cite{Gribov}, calculations involving 
Dyson-Schwinger equations (DSE) for gluon and ghost propagators 
and vertices \cite{DSE1,DSE2}, stochastic quantization \cite{STQ},
functional renormalization group equations \cite{FRG},
and by 
lattice calculations \cite{lattice1,lattice2}.
Most of these calculations suggest that $\A(Q^2)$ remains finite as $Q^2 \to +0$,
i.e., that it has an infrared (IR) fixed point. Furthermore, AdS/CFT
correspondence modified by a dilaton backgound \cite{AdS}
also suggests an IR fixed point.

If spacelike observables $d(Q^2)$ are considered to be functions
of $\A(\kappa Q^2)$ (with $\kappa \sim 1$), then $\A(Q^2)$ 
[$\A(\kappa Q^2)$]  should reflect the
aforementioned analyticity properties of $d(Q^2)$. It is interesting
that imposition of such analyticity properties on $\A(Q^2)$ almost always 
leads to an IR fixed point for $\A(Q^2)$ as well 
\cite{ShS,MS,Sh1Sh2,KS,BMS1,BMS2,BMS3,Webber,1danQCD,2danQCD}. 
This is also true for the perturbation theory in the confining
QCD background in the large-$N_c$ limit \cite{BPT}.

One may also ask whether there exists a renormalization scheme in
which a purely perturbative coupling $a(Q^2)$ is an
analytic function of $Q^2$ in the mentioned sense. In Refs.~\cite{panQCD}
it was shown that it is difficult to construct such models.
Namely, in order to have analyticity of $a(Q^2)$ and 
simultaneously the reproduction of the measured value of the
effective charge for the
semihadronic $\tau$ decay ratio $r_{\tau} \approx 0.20$ ($V+A$ channel),
the required schemes are such that they result in power series 
for observables such that, after a few finite terms, the further
terms appear to be uncontrollably large.

On the other hand, there exist
acceptable analytic models of $\A(Q^2)$ which practically
merge with the perturbative coupling (in the same scheme)
at high $|Q^2| > \Lambda^2$, i.e., $\A(Q^2) - a(Q^2) \sim (\Lambda^2/Q^2)^N$ 
(where $\Lambda^2 \sim 1 \ {\rm GeV}^2$, with $N$ large, e.g., $N=4,5$), 
cf.~Refs.~\cite{Alekseev,Webber,1danQCD,2danQCD}. 
For example, the model 
\cite{2danQCD} has $N=5$ and it reproduces the correct value of $r_{\tau}$.
In such models, due to large $N$ 
the Operator Product Expansion (OPE) approach can be used and interpreted
in the same way as in pQCD (pQCD+OPE), cf.~Ref.~\cite{anOPE}.
Nonetheless, at low energies $|Q^2| < 1 \ {\rm GeV}^2$ the nonperturbative
contributions become appreciable, the theory differs there
appreciably from pQCD.

The coupling $\A(Q^2)$ in some of the models with IR fixed point may have 
Landau singularities within the complex $Q^2$-plane outside the negative
$Q^2$-semiaxis, such as, e.g., the model of Ref.~\cite{StevensonMatt} 
(cf.~the comments on that coupling in Ref.~\cite{panQCD}). However,
in general, it is reasonable to assume that in most of the models with
IR fixed point the analyticity requirement is fulfilled as well, or can be
made fulfilled.

We will present several frameworks with IR fixed point.
In such frameworks it is usually assumed in the literature that
the series in powers of $a(Q^2)$ for physical quantities can be
used unchanged, with the replacements $a(Q^2)^n \mapsto \A(Q^2)^n$,
where $a(Q^2)$ and $\A(Q^2)$ are in the same renormalization scheme
in the perturbative sense.
We will argue that this assumption is 
not correct in IR fixed point scenarios, 
as it leads in general to increasingly
stronger renormalization scale dependence of the result when
the number of terms in the truncated series increases. 
Further, such a series shows tendency to strong divergence,
especially for low-momenta physical quantities, 
partly as a consequence
of the fast growth of the coefficients of the series due to renormalons
(cf.~Ref.~\cite{Beneke:1998ui} and references therein).

In Sec.~\ref{IRmodels} we present several frameworks with IR fixed point.
In Sec.~\ref{problems} we present a construction method
for a nonpower series, in terms of the logarithmic derivatives
$\tA_n(Q^2) \propto d^{n-1} \A(Q^2)/d (\ln Q^2)^{n-1}$, and argue that 
the correct approach for the evaluation of spacelike
QCD physical quantities in the frameworks with IR fixed point 
is in terms of $\tA_n(Q^2)$ and not $\A(Q^2)^n$.
In Sec.~\ref{numev} we present numerical evidence for this,
using as a test case a specific spacelike
physical quantity (massless Adler function) in the
leading-$\beta_0$ (LB) approximation to very high orders, 
and in the full case (``LB+beyondLB'') to the available orders.
We apply evaluations in the usual pQCD (where the running coupling has
unphysical/Landau singularities) and in three chosen scenarios 
with IR fixed point. Specifically, in Sec.~\ref{subs:RScl} 
we present the renormalization scale dependence of various evaluations 
in the various scenarios, and in Sec.~\ref{subs:conv} the
convergence/divergence properties of such evaluations when the
truncation order $N$ increases.
In Sec.~\ref{subs:conv} we apply, in addition,
a resummed version of the logarithmic derivatives approach, 
namely a resummation based on a generalization
of the diagonal Pad\'e resummation. We show that the latter
method is superior to all others in the IR fixed point scenarios.
In Sec.~\ref{timelike} we then argue that for the
timelike physical quantities $\Gamma(s)$ ($s=-Q^2 >0$) 
the evaluation should proceed via the integral transformations 
which relate them with the corresponding spacelike quantities,
where the latter are evaluated with the mentioned approaches.
In Sec.~\ref{summ} we summarize the presented results.

\section{IR fixed point scenarios}
\label{IRmodels}

The simplest case of freezing comes from the use of the the one-loop
perturbative coupling with the replacement $Q^2 \mapsto Q^2+m^2$ where
$m$ is a constant mass (of the order of meson masses) 
\be
\A^{(m)}(Q^2) = 
\frac{1}{\beta_0 \ln \left( \frac{Q^2 + m^2}{\Lambda^2} \right) }
 ,
\label{amrho}
\ee
where $\beta_0= (1/4)(11 - 2 N_f/3)$. It was obtained in 
Ref.~\cite{Simonov} as a consequence of the use of 
nonperturbative QCD background, and is $m \sim 1$ GeV.
It was also used in Refs.~\cite{BKS,KKSh} for an analysis of
structure functions (with $m=m_{\rho} \approx 0.8$ GeV). Similar
construction was made in Ref.~\cite{Shirkovmass}.
This coupling is analytic, in the sense that it has singularities 
in the complex $Q^2$-plane on the negative semiaxis only:
a pole at $Q^2=\Lambda^2 - m^2$ ($<0$), and a cut at $Q^2<-m^2$.
At $Q^2 \to 0$ the coupling freezes at the positive
value $[ \beta_0 \ln(m^2/\Lambda^2)]^{-1}$. 
At large $|Q^2| > \Lambda^2$ it tends to one-loop pQCD coupling and
differs from it by
\be
\A^{(m)}(Q^2) - a^{(1-\ell.)}(Q^2)
\sim \frac{m^2}{Q^2 \ln^2(Q^2/\Lambda^2)} \ .
\label{difmrho}
\ee
\begin{figure}[htb] %\unitlength=1mm
\begin{minipage}[b]{.49\linewidth}
\centering\includegraphics[width=75mm]{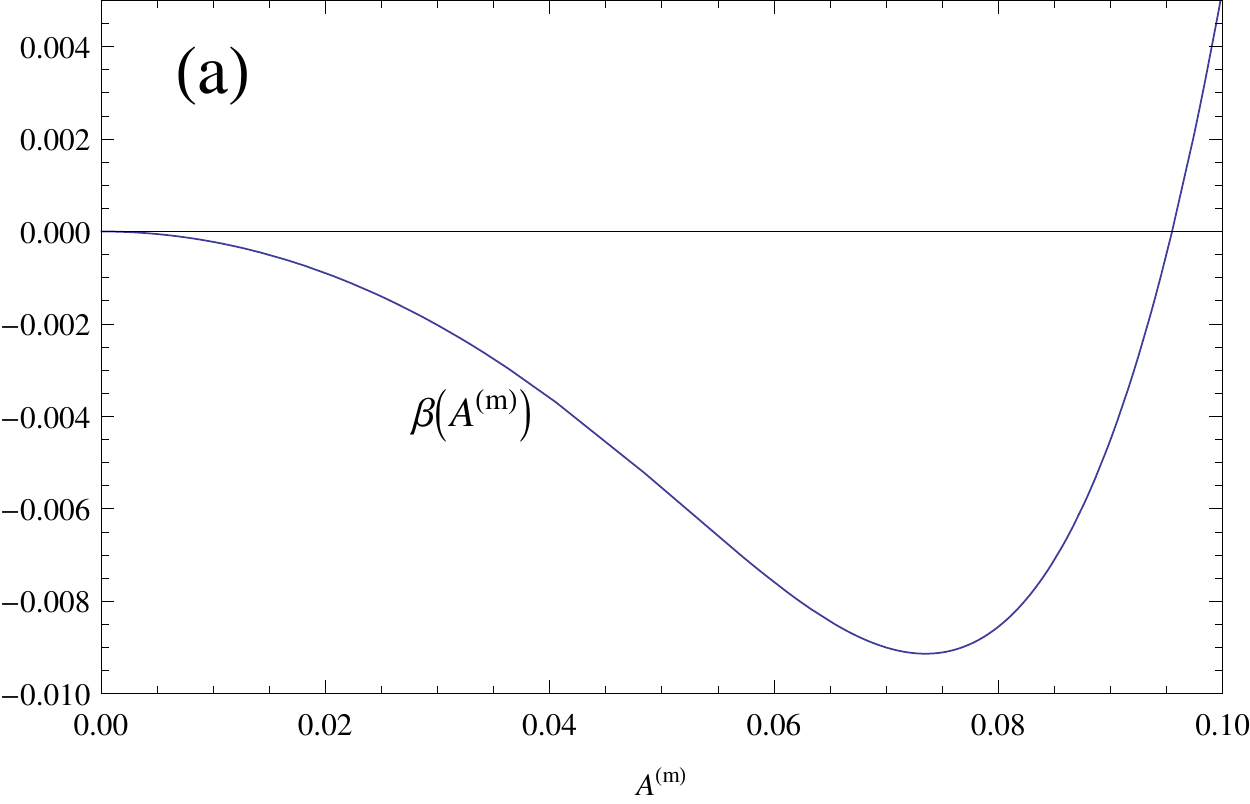}
%\centering\includegraphics[width=75mm]{plbetaAm.pdf}
%centering{\epsfig{file=plbetaAm.eps,width=75mm,angle=0}}
\end{minipage}
\begin{minipage}[b]{.49\linewidth}
\centering\includegraphics[width=75mm]{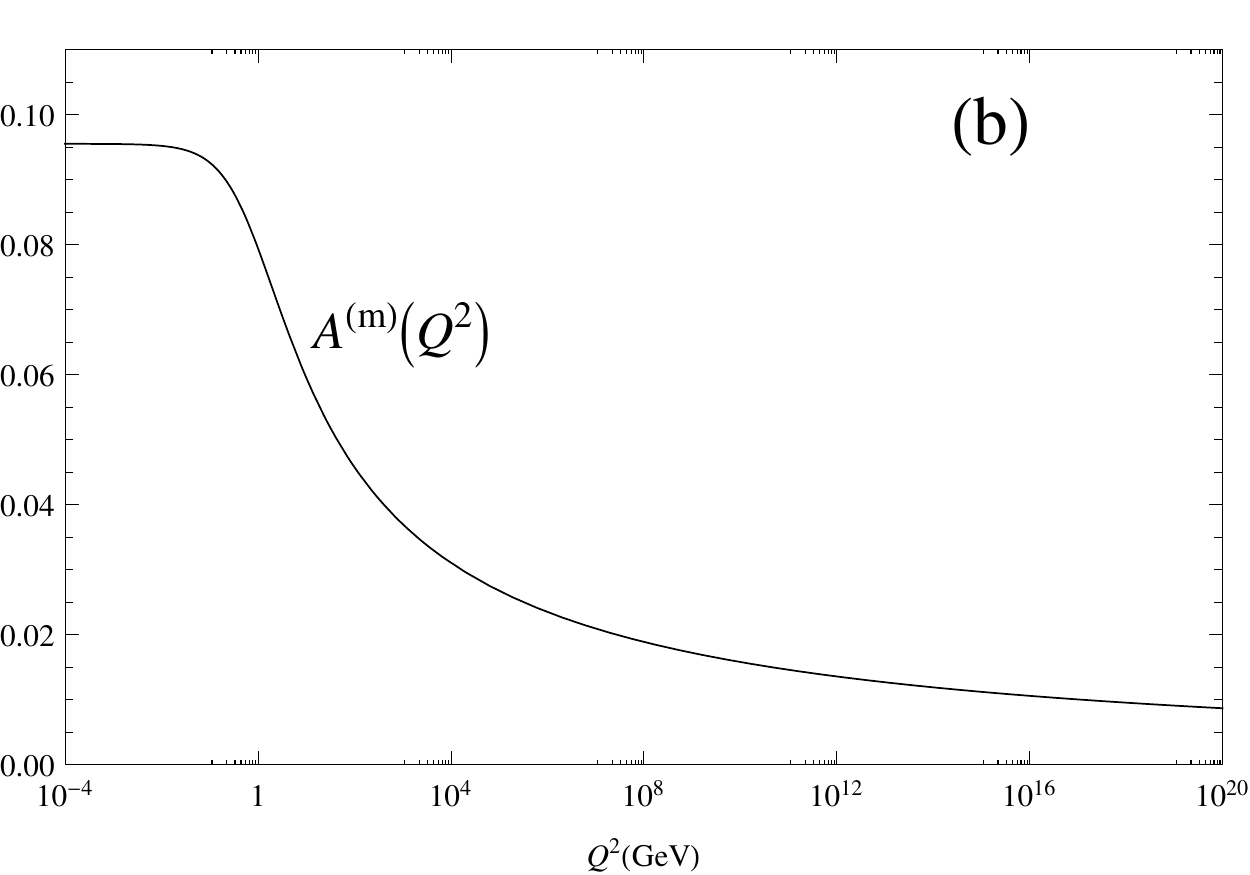}
%\centering\includegraphics[width=75mm]{plAm.pdf}
%centering{\epsfig{file=plAm.eps,width=75mm,angle=0}}
\end{minipage}
\vspace{-0.2cm}
 \caption{(a) Beta function $\beta(\A^{(m)}(Q^2)) = Q^2 d \A^{(m)}(Q^2)/d Q^2$
of the coupling $\A^{(m)}$ of Eq.~(\ref{amrho}); (b) the corresponding
running coupling $\A^{(m)}(Q^2)$ for positive $Q^2$. The values of
the parameters are: $m=0.8$ GeV, $N_f=3$ ($\beta_0=9/4$),
$\Lambda \approx 0.078$ GeV.} 
\label{betAmAm}
 \end{figure}
In Figs.~\ref{betAmAm}(a), (b) we present the beta function 
$\beta(\A^{(m)}(Q^2)) = Q^2 d \A^{(m)}(Q^2)/d Q^2$, and the running
coupling $\A^{(m)}(Q^2)$ at positive $Q^2$, where we chose $N_f=3$, 
$m=0.8$ GeV and $\Lambda$ such that $\A^{(m)}(0) = 0.3/\pi$ ($\Lambda
\approx 0.078$ GeV). 
These curves are qualitatively representative of any IR fixed
point scenario: the freezing of the running coupling
$\A^{(m)}(Q^2)$ at low $Q^2$ (where $Q^2$ is on logarithmic scale), 
and the beta function
achieves zero at $\A^{(m)} = \A^{(m)}(0)$ [$=0.3/\pi \approx 0.0955$ in this 
specific case]. We note that the beta function in this case is
\be
\beta(\A^{(m)}) = - \beta_0 (\A^{(m)})^2 \left[ 1 - \frac{m^2}{\Lambda^2}
\exp \left( - \frac{1}{\beta_0 \A^{(m)}} \right) \right] \ ,
\label{betaAm}
\ee
which is a function of $\A^{(m)}$ nonanalytic at $\A^{(m)}=0$,
implying that this scenario is not of the pQCD-type since
the beta function has a nonperturbative contribution
[$\sim \exp(-1/\beta_0/\A^{(m)})$].

A range of models with similar running of the coupling is suggested
by extensive analyses of the Dyson-Schwinger equations for
the gluon and ghost propagators and vertices \cite{DSE1,DSE2}
and by other functional methods \cite{STQ,FRG}.

At higher $|Q^2|$ ($ > \Lambda^2$), when going beyond the one-loop level,
the multiplicative renormalizability suggests that
the replacement $Q^2 \mapsto (Q^2 + \rho m(Q^2)^2)$ should be made
in the perturbative coupling \cite{Luna}
(cf.~Refs.~\cite{Shirkovmass,BadKuz} when $m$ is constant)
\be
\A^{\rm (DS,n-\ell.)}(Q^2) \approx a^{(n-\ell.)}(Q^2+\rho m(Q^2)^2) \ .
\label{DS2l}
\ee
The dynamical mass $m(Q^2)$ of the DSE-approaches
introduces nonperturbative effects which are felt at $|Q^2| > \Lambda^2$ as
\be
\A^{\rm (DS)}(Q^2) - a(Q^2) \sim \frac{m(Q^2)^2}{Q^2 \ln^2(Q^2/\Lambda^2)} \ .
\label{difDS}
\ee

Coupling with IR fixed point is suggested also by AdS/CFT
correspondence modified by a (positive-sign) dilaton background
\cite{AdS}
\bea
\A^{\rm (AdS mod.)}(Q^2) = \A^{\rm (AdS)}(Q^2) g_{+}(Q^2) +
a^{\rm (fit)}(Q^2) g_{-}(Q^2) \ ,
\label{aAdSmod}
\eea
where at low $Q < 0.8$ GeV predominates the AdS-part
\be
\A^{\rm (AdS)}(Q^2) = \A^{\rm (AdS)}(0) e^{-Q^2/(4 \kappa^2)} \ ,
\label{aAdS}
\ee
with $\kappa = 0.54$ GeV; and $\A^{\rm (AdS)}(0)=1$ is the
IR fixed point in $g_1$ (Bjorken sum rule)
effective charge scheme.\footnote{
It turns out that the same coupling can be obtained also in 
the negative-sign dilaton scenario; the five-dimensional coupling
is defined in both cases as $g_5^{-2}(z) = e^{\phi(z)} g_5^{-2}$ where 
$\phi(z)=\kappa^2 z^2$; the sign of the dilaton affects neither 
the running coupling nor the mass spectrum, but 
becomes important for the calculation of the bulk-to-boundary propagator
in the AdS space, Ref.~\cite{Lyubovitskij}.}    
On the other hand, $a^{\rm (fit)}(Q^2)$ is
obtained by fit to the data for $Q > 0.8$ GeV. $g_{\pm}(Q^2)$ are 
smeared step functions, e.g., $g_{\pm}(Q^2) = 1/(1 + e^{\pm(Q^2-Q_0^2)/\tau^2})$
with $Q_0=0.8$ GeV and $\tau = \kappa$. At large $|Q^2| > \kappa^2$ the difference
between this coupling and the perturbative coupling is
very small
\be 
\A^{\rm (AdS mod.)}(Q^2) - a(Q^2) \sim 
\frac{e^{-Q^2/\kappa^2}}{\ln (Q^2/\Lambda^2)}
\qquad (|Q^2| \gg \kappa^2) \ .
\label{difAdS}
\ee

\begin{figure}[htb]
\includegraphics[width=120mm]{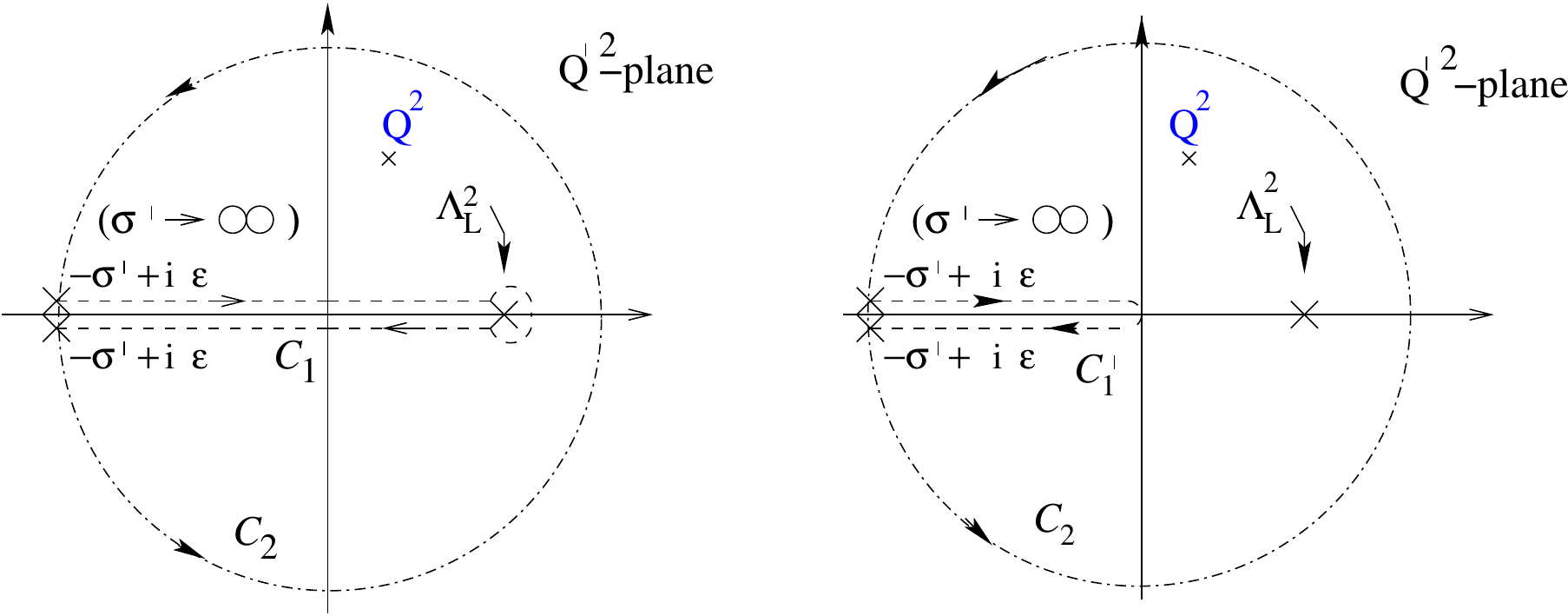}
\vspace{-0.4cm}
\caption{Left-hand Figure: the integration path for the 
integrand $a_{\rm pt}(Q'^2)/(Q'^2 - Q^2)$ leading to the 
dispersion relation (\ref{aptdisp}) for  $a_{\rm pt}(Q^2)$. 
Right-hand Figure: the integration part for the same integrand, 
leading to the dispersion relation (\ref{MAAnudisp}) for the APT 
coupling $\A^{\rm (APT)}(Q^2)$.}
\label{intpath}
\end{figure}

Another case is the Analytic Perturbation Theory (APT) coupling
\cite{ShS,MS,Sh1Sh2}, which is obtained by ``minimally'' analytizing the
perturbative ($n$-loop) coupling $a(Q^2)$. The
construction of the APT coupling is the following.
The pQCD coupling has singularities on the semiaxis $Q^2 < \Lambda_{\rm L}^2$,
where the (Landau) cut $0 < Q^2 < \Lambda_{\rm L}^2$ 
starts at the branching point $\Lambda_{\rm L}^2$, 
and is unphysical in the aforementioned
sense. Application of the Cauchy theorem to the function
$a(Q^{'2})/(Q^{'2}-Q^2)$ to an appropriate closed contour 
(avoiding the cuts) in the complex $Q^{'2}$-plane, leads to
the following dispersion relation for $a(Q^2)$
\begin{equation}
a(Q^2) = \frac{1}{\pi} \int_{\sigma= - {\Lambda_{\rm L}}^2 - \eta}^{\infty}
\frac{d \sigma {\rho^{\rm {(pt)}}}(\sigma) }{(\sigma + Q^2)},
   \quad (\eta \to +0),
\label{aptdisp}
\end{equation}
where ${\rho^{\rm {(pt)}}}(\sigma)$ is the pQCD 
discontinuity function of $a$
along the cut axis:
${\rho^{\rm {(pt)}}}(\sigma)= {\rm Im} a(-\sigma - i \epsilon)$.
The APT procedure consists in the elimination, in the above integral,
of the contributions of the Landau cut $0 < (-\sigma) \leq \Lambda^2$, leading to
the APT analytic analog of $a$ (see Fig.~\ref{intpath})
\begin{equation}
\A^{\rm (APT)}(Q^2) = \frac{1}{\pi} \int_{\sigma= 0}^{\infty}
\frac{d \sigma {\rho^{\rm {(pt)}}}(\sigma) }{(\sigma + Q^2)} \ .
\label{MAA1disp}
\end{equation}
The APT analogs of powers $a^{\nu}$ ($\nu$ a real exponent) is
obtained in the same way
\begin{equation}
{\A}^{\rm {(APT)}}_{\nu}(Q^2) = \frac{1}{\pi} \int_{\sigma= 0}^{\infty}
\frac{d \sigma {\rho^{\rm {(pt)}}_{\nu}}(\sigma) }{(\sigma + Q^2)} \ ,
\label{MAAnudisp}
\end{equation}
where ${\rho^{\rm {(pt)}}_{\nu}}(\sigma) = {\rm Im} a^{\nu}(-\sigma - i \epsilon)$.
The underlying pQCD coupling $a(Q^2)$ can run at any $n$-loop
level and can be in any chosen renormalization scheme; the
corresponding renormalization group equation (RGE) is
\begin{eqnarray}
\frac{\partial a(\ln Q^2; {\beta_2}, \ldots)}
{\partial \ln Q^2} 
& = &
- \sum_{j=0}^{n-1} \beta_{j} \: 
a^{j+2} (\ln Q^2; {\beta_2}, \ldots) ,
\label{pRGE}
\end{eqnarray}
where the first two beta coefficients are universal 
[$\beta_0= (1/4)(11-2 N_f/3)$, $\beta_1=(1/16)(102-38 N_f/3)$],
and the other coefficients $\beta_k$ ($k \geq 2$) characterize
the perturbative renormalization scheme. It turns out that
the APT coupling has IR fixed point: $\A(0)=1/\beta_0$
($=4/9 \approx 0.44$ if $N_f=3$). At one-loop level, it is particularly
simple:
\be
\A^{\rm (APT, 1-\ell.)}(Q^2) = \frac{1}{\beta_0} \left[
\frac{1}{\ln z} - \frac{1}{(z-1)} \right] \, 
\quad (z \equiv Q^2/\Lambda^2) \ .
\label{MAA11l}
\ee
Explicit expressions for $\A_{\nu}^{\rm (APT)}$ at one-loop level also
exist and were constructed and used in Ref.~\cite{BMS1}
\be
\A_{\nu}(Q^2)^{\rm (APT, 1-\ell.)} = \frac{1}{\beta_0^{\nu}}
\left(  \frac{1}{\ln^{\nu}(z)} -
\frac{ {\rm Li}_{-\nu+1}(1/z)}{\Gamma(\nu)} \right) \ ,
\label{MAAnu1l}
\ee
where $z \equiv Q^2/\Lambda^2$ and 
${\rm Li}_{-\nu+1}(z)$ is the polylogarithm function of order $-\nu+1$.
Extensions to higher loops were performed via expansions of the
one-loop result \cite{BMS2,BMS3} [Fractional APT (FAPT)]. For a
review of FAPT, see Refs.~\cite{Bakulev}, and mathematical packages 
for numerical calculation are given in Refs.~\cite{BK}.   

Another analytic model, based on the minimal analytization
of the function $d \ln a(Q^2)/d \ln Q^2$, Refs.~\cite{Nesterenko1}, 
leads to a coupling with no freezing in the IR.

It turns out that the APT coupling differs from the pQCD coupling
by terms $\sim (\Lambda^2/Q^2)$ at large $|Q^2| > \Lambda^2$
\be
\A^{\rm (APT)}(Q^2) - a(Q^2) \sim 
\left( \frac{\Lambda^2}{Q^2} \right)^1 \ ,
\label{difAPT}
\ee
which may be appreciable even at high energies. An extension of
the APT coupling at one-loop, such that the difference between
it and the pQCD coupling is $\sim  (\Lambda^2/Q^2)^p$, was 
proposed by Webber \cite{Webber}
\begin{equation}
\A^{\rm (W, 1-\ell.)}(Q^2) = \frac{1}{{\beta_0}} \left[
\frac{1}{\ln z} + \frac{1}{1-z}\,\frac{z+{b}}{1+{b}}
\left(\frac{1+{c}}{z+{c}}\right)^{\!{p}}\,
\right], 
\label{AW}
\end{equation}
where $z \equiv Q^2/\Lambda^2$ and specific values of parameters were
chosen such that the model gives good agreement with a range of data on
power corrections: $b=1/4$, $c=4$, and $p=4$.
The coupling has IR fixed point, 
$\A^{\rm (W, 1-\ell.)}(0) = 1/(2 {\beta_0}) \approx 0.22$. In this model,
the difference from the pQCD coupling is
\be
 \A^{\rm (W, 1-\ell.)}(Q^2) - a^{\rm (1-\ell.)}(Q^2)
\sim \left( \frac{\Lambda^2}{Q^2} \right)^4 \ .
\label{difW}
\ee

A somewhat related is the model of Alekseev \cite{Alekseev} for
the coupling $\A(Q^2)$ (called synthetic coupling), 
which is a modification of the APT model
$\A$ (at any loop-level) 
\begin{equation}
\A^{\rm (Al.)}(Q^2)= \A^{\rm (APT)}(Q^2)
+\frac{1}{\beta_0}\left[ \frac{{c}{\Lambda}^2}{Q^2}-
\frac{{d} \; {\Lambda}^2}
{Q^2+{m_g}^2}\right], 
\label{A1Al}
\end{equation}
where the three parameters $c$, $d$ and the effective gluon mass $m_g$ were
determined by the requirement
\be
\A^{\rm (Al.)}(Q^2) - a(Q^2) \sim 
\left( \frac{\Lambda^2}{Q^2} \right)^3 \ 
\label{difAl}
\ee
and by the string tension parameter in the IR.
However, in this case, there is no IR fixed point, due to the
term $\sim 1/Q^2$ in the constructed coupling.

Yet another approach is based on the general dispersive relation 
for analytic couplings,
\begin{equation}
\A(Q^2) = \frac{1}{\pi} \int_{\sigma= 0}^{\infty}
\frac{d \sigma {\rho}(\sigma) }{(\sigma + Q^2)} \ ,
\label{A1disp}
\end{equation}
where  ${\rho}$ is the discontinuity function of $\A$:
${\rho}(\sigma) = {\rm Im} \A(-\sigma - i \epsilon)$. 
In Refs.~\cite{1danQCD,2danQCD} this discontinuity function
was approximated at high momenta $\sigma \geq M_0^2$ ($\agt 1 \ {\rm GeV}^2$)
by its pQCD analog $\rho^{\rm (pt)}(\sigma)={\rm Im} a(-\sigma - i \epsilon)$;
in the unknown low-energy regime, $0 < \sigma < M_0^2$
it was approximated by either one (Ref.~\cite{1danQCD}) 
or two delta functions (Ref.~\cite{2danQCD}) 
\bes
\bea
\rho(\sigma)^{(1 \delta)}(\sigma) & = & \pi F_1^2 \delta(\sigma - M_1^2)
+ \Theta(\sigma-M_0^2) \rho^{\rm (pt)}(\sigma) \ ,
\label{rho1d}
\\
\rho(\sigma)^{(2 \delta)}(\sigma) & = & \pi F_1^2 \delta(\sigma - M_1^2)
+ \pi F_2^2 \delta(\sigma - M_2^2) + \Theta(\sigma-M_0^2) \rho^{\rm (pt)}(\sigma) \ .
\label{rho2d}
\eea
\ees
The parameters
of the delta functions and the pQCD-onset scale $M_0$ were
adjusted so that the correct value of the semihadronic tau decay
ratio $r_{\tau} \approx 0.20$ ($V+A$ channel) was reproduced and that
the difference from the pQCD coupling at high $|Q^2| > \Lambda^2$
is as strongly suppressed as possible
\bes
\bea
\A^{(1 \delta)}(Q^2) &=&  \frac{F_1^2}{Q^2 + M_1^2}
+ \frac{1}{\pi} \int_{M_0^2}^{\infty} d \sigma \; 
\frac{\rho^{\rm (pt)}(\sigma)}{(Q^2+\sigma)} \ ,
\label{1dA1}
\\
\A^{(2 \delta)}(Q^2) &=&  \frac{F_1^2}{Q^2 + M_1^2} +
\frac{F_2^2}{Q^2 + M_2^2}
+ \frac{1}{\pi} \int_{M_0^2}^{\infty} d \sigma \; 
\frac{\rho^{\rm (pt)}(\sigma)}{(Q^2+\sigma)} \ .
\label{2dA1}
\eea
\ees
The resulting deviations from pQCD at high $|Q^2| > \Lambda^2$ are
\bes
\bea
\A^{(1 \delta)}(Q^2) - a(Q^2) & \sim & 
\left( \frac{\Lambda^2}{Q^2} \right)^3 \ ,
\label{dif1d}
\\
\A^{(2 \delta)}(Q^2) - a(Q^2) & \sim & 
\left( \frac{\Lambda^2}{Q^2} \right)^5 \ .
\label{dif2d} 
\eea
\ees
The suppression of deviation from pQCD coupling, at high $|Q^2|$,
may be regarded as preferred because then OPE can be used and
interpreted in such models in the same way as OPE in pQCD --
that the higher dimensional nonperturbative terms $\sim 1/(Q^2)^N$ 
($N \leq N_{\rm cr.}$, with $N_{\rm cr.}=4$ in $2 \delta$ model) 
have purely IR origin. In APT, in view of the significant 
difference (\ref{difAPT}), such interpretation is not possible, 
and part of the nonperturbative terms $\sim 1/(Q^2)^N$ ($N=1,2,\ldots$)
comes from the leading-twist contribution and has UV origin.
 
Both models ($1 \delta$, $2 \delta$) have IR fixed point, with $\A(0) \leq 1$.

\section{Problems with power series for spacelike 
physical quantities in IR fixed point scenarios, and solution}
\label{problems}

If we regard a spacelike physical quantity $d(Q^2)$, 
such as current correlators (Adler function, etc.) or 
structure function sum rules, the usual evaluation in pQCD
is via the power series
\bea
d(Q^2)_{\rm pt} &=& a(\kappa Q^2) +
\sum_{n=1}^{\infty} d_n(\kappa) \; a(\kappa Q^2)^{n+1} \ ,
\label{Dpt}
\eea    
where $\mu^2 = \kappa Q^2$ is a renormalization scale ($\kappa \sim 1$).
The dependence on other renormalization scheme parameters ($\beta_2, \beta_3,
\ldots$) has been suppressed in the notation.
Unless this series is the leading-$\beta_0$ resummation or some other
partial resummation, the series is known only up to certain order 
$\sim a^N$ (usually $N=3$ or $4$)
\be
d(Q^2; \kappa)_{\rm pt}^{[N]} = a(\kappa Q^2) + 
\sum_{j=1}^{N - 1} d_j(\kappa) \; a(\kappa Q^2)^{j+1} \ .
\label{Dpttr}
\ee
As a consequence of truncation, the truncated series has unphysical
dependence on the renormalization scale (RS) parameter $\kappa$.
However, the more terms are included, the weaker is the RS dependence (at high $|Q^2|$) generally
\be
\frac{ \partial d_{\rm pt}^{[N]}}{\partial \ln \kappa}
= K_N a(\kappa Q^2)^{N+1} +  K_{N+1} a(\kappa Q^2)^{N+2} + \cdots 
\; \sim a^{N+1} \ ,
\label{RSdepDpt}
\ee 
where $K_N, K_{N+1},\ldots$ are specific coefficients determined by the
original truncated series coefficients $d_n(\kappa)$ 
($n=1,\ldots,N-1$).\footnote{The RS dependence at low $\kappa Q^2$ is,
however, strengthened by the large value of $a(\kappa Q^2)$, as
seen from Eq.~(\ref{RSdepDpt}).}
If we now have a model where the coupling $\A(Q^2)$
has nonperturbative contributions (such as any of the aforementiond
IR fixed point scenarios), we have
\be
\A(Q^2) - a(Q^2) = T_{\rm NP}(Q^2) \ ,
\label{difNP}
\ee
where the term $T_{\rm NP}(Q^2)$ is nonperturbative,
i.e., at $|Q^2| > \Lambda^2$ 
it is a function of $a(Q^2)$, $T_{\rm NP}(Q^2) = F(a(Q^2))$,
which is nonanalytic at $a=0$. For example, 
\bes
\label{NP}
\bea
T_{\rm NP}(Q^2) &\sim& \left( \frac{\Lambda^2}{Q^2} \right)^n \approx 
\exp \left[- \frac{n}{\beta_0 a(Q^2)} \right] \ ,
\\
T_{\rm NP}(Q^2) &\sim& \exp\left( - \frac{Q^2}{K^2} \right) \sim 
\exp \left[ -\left( \frac{\Lambda^2}{K^2}\right) 
e^{1/\beta_0 a(Q^2)} \right] \ .
\eea
\ees
If applying now the power series (\ref{Dpt}) in the 
IR fixed point scenarios
\be
d(Q^2; \kappa)_{\rm pt, \A}^{[N]} = \A(\kappa Q^2) + 
\sum_{j=1}^{N - 1} d_j(\kappa) \; \A(\kappa Q^2)^{j+1} \ ,
\label{DptAtr}
\ee
the inclusion of more terms in this power series tends to make
the result increasingly more RS-dependent\footnote{
On the other hand, the couplings at low momenta are in general smaller
than in the underlying pQCD, and this effects tends to make the
RS dependence of the truncated power series smaller than in pQCD.} 
or the RS dependence becomes more erratic,
due to the inclusion of
the effects of the NP terms $\sim T_{\rm NP}(\kappa Q^2)^k \A(\kappa Q^2)^{m}$.
This can be numerically verified.
These aspects are also reflected in the fact that
the beta function in all the aforedescribed IR fixed point scenarios,
$\beta(\A(Q^2)) \equiv \partial \A(Q^2)/\partial \ln Q^2$, cannot be presented 
fully with a power expansion in $\A$, i.e., $\beta(\A)$ contains also terms
which are nonanalytic in $\A$.

All this suggests that the analog of
the power $a^n$ is not $\A^n$, but rather a nonpower 
expression $\A_n$. Within the context of APT \cite{ShS}, this has been
noted by the authors of APT, and the construction Eq.~(\ref{MAAnudisp})
really gives $\A_{\nu}^{\rm (APT)} \not=  (A_1^{\rm (APT)})^{\nu}$. However,
in general models with finite $\A(0)$, the APT-type of
construction cannot be made since it uses only the
pQCD couplings (and their discontinuities).

It turns out that the construction of $\A_n$, the analog of 
$a^n$, can be made in such general frameworks with IR fixed point,
via a detour by construction of logarithmic derivatives, 
\cite{CV1,CV2,panQCD}. In pQCD these are
\be
{\ta}_{n+1}(Q^2)
\equiv \frac{(-1)^{n}}{\beta_0^{n} n!}
\frac{ \partial^n a(Q^2)}{\partial (\ln Q^2)^n} \ , 
\qquad (n=1,2,\ldots) \ .
\label{tan}
\ee
We note that ${\ta}_{n+1}(Q^2) = a(Q^2)^{n+1} + {\cal O}(a^{n+2})$
by RGE $\partial a(Q^2)/\partial \ln Q^2 = \beta(a(Q^2))$, where
beta function $\beta(a)$ has the pQCD expansion as given in
Eq.~(\ref{pRGE}).
The analytization is a linear operation. Therefore
\be
(a(Q^2))_{\rm an} = \A(Q^2) \ \Rightarrow \
\left( \frac{\partial a(Q^2)}{\partial \ln Q^2} \right)_{\rm an} 
= \frac{\partial \A(Q^2)}{\partial \ln Q^2} \ .
\label{linan}
\ee
Therefore, in general
\be
\left( {\ta}_{n+1}(Q^2) \right)_{\rm an} = \tA_{n+1}(Q^2) \ ,
\label{tatA}
\ee
where 
\be
\tA_{n+1}(Q^2)
\equiv \frac{(-1)^n}{\beta_0^n n!}
\frac{ \partial^n \A(Q^2)}{\partial (\ln Q^2)^n} \ .
\qquad (n=1,2,\ldots) \ .
\label{tAn}
\ee
In virtually all IR fixed point (analytic) models
we have: $|\A(Q^2)| > |\tA_2(Q^2)| > |\tA_3(Q^2)| > \cdots$
for any $Q^2$ (not just when $|Q^2|$ is large). This is
an empirical observation which could possibly be proven
under specific conditions.

The basic relation (\ref{tatA}) then requires reexpression
of the power series (\ref{Dpt}) as a series in logarithmic
derivatives ${\ta}_{n+1}(Q^2)$ (``modified'' perturbation series, mpt)
\bea
d(Q^2)_{\rm mpt} &=& a(\kappa Q^2) + 
\sum_{n=1}^{\infty} {\td}_n(\kappa) \; {\ta}_{n+1}(\kappa Q^2) \ .
\label{Dmpt}
\eea
This leads, after the analytization  (\ref{tatA}) term-by-term,
to the ``modified'' analytic (man) series 
\bea
d(Q^2)_{\rm man} &=& \A(\kappa Q^2) + 
\sum_{n=1}^{\infty} {\td}_n(\kappa) \; \tA_{n+1}(\kappa Q^2) \ .
\label{Dman}
\eea 
This is the basic expression for evaluation of $d(Q^2)$ in
IR fixed point scenarios.
Incidentally, also the mpt truncated series 
\be
d(Q^2; \kappa)_{\rm mpt}^{[N]} = a(\kappa Q^2) + 
\sum_{j=1}^{N - 1} \td_j(\kappa) \; \ta_{j+1}(\kappa Q^2) \ ,
\label{Dmpttr}
\ee
has RS dependence due to truncation, similar to the dependence 
(\ref{RSdepDpt}) of the truncated pt series, but even simpler
\be
\frac{ \partial d_{\rm mpt}^{[N]}}{\partial \ln \kappa}
= -\beta_0 N \td_{N-1}(\kappa) \ta_{N+1}(\kappa Q^2) \ .
\label{RSdepDmpt}
\ee 
The truncated modified analytic series is
\bea
d(Q^2;\kappa)^{[N]}_{\rm man} &=& \A(\kappa Q^2) + 
\sum_{j=1}^{N-1} {\td}_j(\kappa) \; \tA_{j+1}(\kappa Q^2) \ .
\label{Dmantr}
\eea 

The mpt series (\ref{Dmpt}) is just a reorganization of the original
perturbation (pt) series (\ref{Dpt}), so it is also RS-independent.
In conjunction with the recurrence relation 
$\partial \ta_n(\kappa Q^2)/\partial \ln \kappa = - \beta_0 n \ta_{n+1}(\kappa Q^2)$
which follows from the definition (\ref{tan}),
we obtain simple differential relations between
$\td_n(\kappa)$:
\be
\frac{d}{d \ln \kappa} \td_n(\kappa) = n \beta_0 \td_{n-1} (\kappa) 
\qquad (n=1,2,\ldots) \ .
\label{tdndiff}
\ee
($d_0(\kappa)={\td}_0(\kappa) = 1$ by definition). Integrating them, the
renormalization scale dependence of the coefficients $\td_n$ is
particularly simple
\be
{\td}_n(\kappa) = {\td}_n(1) + \sum_{k=1}^n 
\left(
\begin{array}{c}
n \\
k
\end{array}
\right)
\ \beta_0^k \ \ln^k ( \kappa ) {\td}_{n-k}(1) \ .
\label{tdnmu}
\ee
($\kappa \equiv \mu^2/Q^2$; $d_0 = \td_0=1$).
The coefficients $\td_n(\kappa)$ are obtained from $d_k(\kappa)$'s
($k \leq n$) in the following way. First we express
the logarithmic derivatives $\ta_{n+1}$ in terms of the powers
$a^{k+1}$, at a given scale $Q^2$ or $\mu^2 = \kappa Q^2$, 
using the RGE relations in pQCD for these powers
[RGE (\ref{pRGE}) and its derivatives]
\bes
\label{tas}
\bea
\ta_{2} &=& a^2 + c_1 a^3 + c_2 a^4 + \cdots \ ,
\label{ta2}
\\
\ta_{3} &=& a^3 + \frac{5}{2} c_1 a^4  + \cdots \ ,
\qquad
\ta_{4} = a^4 +   \cdots \ ,  
\qquad {\rm etc.} \ ,
\label{ta3ta4}
\eea
\ees 
where we use the notation $c_j \equiv \beta_j/\beta_0$.
We now invert them
\bes
\label{as}
\bea
a^2 & = & \ta_{2}
- c_1 \ta_{3} 
+ \left( \frac{5}{2} c_1^2 - c_2 \right) \ta_{4} + \cdots \ ,
\label{a2}
\\
a^3 & = & \ta_{3} - \frac{5}{2} c_1 \ta_{4} + \cdots \ ,
\qquad
 a^4  =  \ta_{4}  +  \cdots \ ,
\qquad {\rm etc.}
\label{a3a4}
\eea
\ees
Replacing these relations into the original perturbation
expansion (\ref{Dpt}) for $d(Q^2)$, the
coefficients $\td_n(\kappa)$ of the reorganized (``modified'')
expansions (\ref{Dmpt})-(\ref{Dman}) can be read off
\bes
\label{tds}
\bea
\td_1(\kappa) & = & d_1(\kappa) \ , \qquad
\td_2(\kappa) = d_2(\kappa) - c_1 d_1(\kappa) \ ,
\label{td1td2}
\\
\td_3(\kappa) & = & d_3(\kappa) - \frac{5}{2} c_1 d_2(\kappa)
+ \left( \frac{5}{2} c_1^2 - c_2 \right) d_1(\kappa) \ ,
\qquad {\rm etc.}
\label{td3}
\eea
\ees
Now we perform analytization, Eqs.~(\ref{tatA})-(\ref{tAn}), in relations
(\ref{a2})-(\ref{a3a4}) term-by-term. In this way
we obtain the (IR fixed point) analogs of integer powers $a^n$, 
$\A_n = (a^n)_{\rm an}$ 
\bes
\label{As}
\bea
\A_2 & \equiv & \left( a^2 \right)_{\rm an} 
= \tA_2 - c_1 \tA_3
+ \left( \frac{5}{2} c_1^2 - c_2 \right) \tA_{4} + \cdots \ ,
\label{A2}
\\
\A_3 & \equiv & \left( a^3 \right)_{\rm an} 
=  \tA_3 - \frac{5}{2} c_1 \tA_4 + \cdots \ ,
\quad
 \A_4 \equiv \left( a^4 \right)_{\rm an} 
=  \tA_{4}  +  \cdots \ ,
\qquad {\rm etc.}
\label{A3A4}
\eea
\ees
This allows us to reexpress the ``modified'' analytic series (\ref{Dman}) 
in a form which is in close analogy with the
original perturbation series (\ref{Dpt})
\bea
d(Q^2)_{\rm an} &=& \A(\kappa Q^2) + 
\sum_{n=1}^{\infty} d_n(\kappa) \; \A_{n+1}(\kappa Q^2) \ .
\label{Dan}
\eea 
This series is $\kappa$-independent since it coincides
with the series $d(Q^2)_{\rm man}$
of Eq.~(\ref{Dman}). The truncated series is
\bea
d(Q^2;\kappa)^{[N]}_{\rm an} &=& \A(\kappa Q^2) + 
\sum_{n=1}^{N-1} d_n(\kappa) \; \A_{n+1}(\kappa Q^2) \ .
\label{Dantr}
\eea 
When we truncate the relations (\ref{As}) at $\tA_N$, it is
straightforward to check that the truncated series (\ref{Dantr})
coincides with the truncated series (\ref{Dmantr})
\be
d(Q^2;\kappa)^{[N]}_{\rm an} = d(Q^2;\kappa)^{[N]}_{\rm man} \ .
\label{aneqman}
\ee
The quantities $\tA_{n}$ and $\ta_n$ have the same RS dependence
relations (just interchanging $\tA_{n} \leftrightarrow \ta_n$);
and the quantities $\A_{n}$ and $a^n$ have the same RS-dependence
relations (just interchanging $\A_{n} \leftrightarrow a^n$). This implies that
the structure of the RS-dependence of the
truncated pt and mpt series in pQCD, 
Eqs.~(\ref{RSdepDpt}) and (\ref{RSdepDmpt}), survives in its 
analytic form for the truncated analytic (\ref{Dantr}) and
modified analytic series (\ref{Dmantr}) 
\bes
\label{RSdepan}
\bea
\frac{ \partial d_{\rm an}^{[N]}}{\partial \ln \kappa}
&=& K_N \A_{N+1}(\kappa Q^2) 
+  K_{N+1} \A_{N+2}(\kappa Q^2) + \cdots \ .
\label{RSdepDan}
\\
\frac{ \partial d_{\rm man}^{[N]}}{\partial \ln \kappa}
&=& -\beta_0 N \td_{N-1}(\kappa) \tA_{N+1}(\kappa Q^2) \ ,
\label{RSdepDman}
\eea
\ees
When the truncations in the construction of $\A_n$'s,
Eqs. (\ref{As}), are made at $\tA_N$, we have the coincidence
of the two truncated series, Eq.~(\ref{aneqman}),
and then the right-hand side of Eq.~(\ref{RSdepDan}) can be written
in the simpler form of the right-hand side of Eq.~(\ref{RSdepDman}).

These relations, in conjunction with the aforementioned
hierarchy $|\A(Q^2)| > |\tA_2(Q^2)| > ...$ and hierarchy
$|\A(Q^2)| > |\A_2(Q^2)| > ...$, at all $Q^2$ (not just high $|Q^2|$),
suggest that the truncated analytic series 
$d_{\rm man}^{[N]}(Q^2;\kappa)$, Eq.~(\ref{Dmantr}), and
$d_{\rm an}^{[N]}(Q^2;\kappa)$, Eq.~(\ref{Dantr}), have
in general weaker RS dependence when the number of terms 
increases,\footnote{
The renormalon growth of the coefficients $\td_N$ with increasing
$N$ eventually increases the RS dependence.}
or that the RS dependence is more under control (less erratic) than
in the case of truncated series in powers of $\A$.
 This is true even for low-energy
quantities (i.e., when $|Q^2|$ is low), in contrast to the case of
perturbative truncated series $d(Q^2;\kappa)_{\rm pt}^{[N]}$ and
$d(Q^2;\kappa)_{\rm mpt}^{[N]}$.

Further, the described construction is applicable even in the
scenarios without IR fixed point, as long as the analyticity
of $\A(Q^2)$ is valid in the complex $Q^2$-plane outside the 
semiaxis $Q^2 \leq 0$, e.g, the model of Refs.~\cite{Nesterenko1}.

All the above considerations can be extended in the same spirit 
to the case of the subleading renormalization scheme dependence, i.e.,
dependence on the scheme parameters $c_j = \beta_j/\beta_0$ ($j=2,3,\ldots$).
We refer to Appendix \ref{app1} for a few details about this
aspect.

The construction of $\tA_n$ and $\A_n$ was demonstrated here for integer $n$.
However, for noninteger $n=\nu$ these quantities can also be obtained \cite{GCAK}, via an analytic continuation of the general formulas in $n \mapsto \nu$).
We refer to Appendix \ref{app2} for some of the details of the
construction of $\tA_{\nu}$ and $\A_{\nu}$.

\section{Numerical evidence: the case of Adler function}
\label{numev}

\subsection{Renormalization scale dependence of truncated series}
\label{subs:RScl}

Here we will illustrate the arguments of the
previous Section numerically, in the case of a specific
massless spacelike observable, 
for the truncated power series and
the truncated series in logarithmic derivatives,
within various IR fixed point frameworks.
We will consider the massless Adler function.
The effective charge of the (massless) Adler function is defined as
\be
d_{\rm Adl}(Q^2) = - (2 \pi^2) \frac{d \Pi(Q^2)}{d \ln Q^2} - 1 \ , 
\label{ddef}
\ee
whose pQCD power expansion (pt) is
\be
d_{\rm Adl}(Q^2)_{\rm pt}
= a(Q^2) + d_1 a(Q^2)^2 + \cdots \ ,
\label{dpt}
\ee
and where $\Pi(Q^2) = \Pi_V(Q^2)+\Pi_A(Q^2)$ ($= 2 \Pi_V(Q^2)$, in the massless case)
is the correlator of the nonstrange charged hadronic currents
\bea
\Pi^V_{\mu\nu}(q) & = & i \int  d^4 x \; \exp( i q \cdot x) 
\langle T  V_{\mu}(x) V_{\nu}(0)^{\dagger} \rangle
=  (q_{\mu} q_{\nu} - g_{\mu \nu} q^2) \Pi_V(Q^2) \ ,
\label{PiV}
\eea
where: $V_{\mu} = {\overline u} \gamma_{\mu} d$.
The leading-$\beta_0$ (LB) part of this spacelike quantity
is known to all orders
\bes
\label{DLB}
\bea
d_{\rm Adl}^{\rm (LB)}(Q^2)_{\rm mpt} & = &   
\int_0^{\infty} \frac{dt}{t} \; F_d(t) a(t Q^2 e^{{{\cal C}}})
\label{DLBint}
\\
& = &
a(Q^2) + \td_{1}^{\rm (LB)} {\ta}_{2}(Q^2) +
\cdots +  \td_{n}^{\rm (LB)} {\ta}_{n+1}(Q^2) + \cdots
\label{DLBmpt}
\\
& = & a(Q^2) + d_{1}^{\rm (LB)} a(Q^2)^2 + 
\cdots d_{n}^{\rm (LB)} a(Q^2)^{n+1} + \cdots 
\label{DLBpt}
\eea
\ees
where $F_d(t)$ is the distribution function of the Adler function.
It was obtained in Ref.~\cite{Neubert} on the basis of the
LB expansion coefficients 
$\td_{n}^{\rm (LB)} \equiv \td_{n,n} \beta_0^n$ 
obtained from the
LB Borel transform of Refs.~\cite{Broad1,Broad2} 
(cf.~also \cite{Ben}). 
The value of $a(t Q^2 e^{\cal C})$ is independent of the
scaling convention ($\Lambda$ definition). Here we take the $\MSbar$
scaling convention: ${{\cal C}}=-5/3$.
We refer to Appendix \ref{appFd} for details on the formulas (\ref{DLB}).
The couplings $a$ and $\ta_{n+1}$ in Eqs.~(\ref{DLB}) are 
considered here to be general ($N$-loop) couplings, and in the
IR fixed point frameworks they are replaced by $\A$ and $\tA_{n+1}$,
respectively.

We stress that what was obtained in Refs.~\cite{Broad1,Broad2}
are the coefficients $\td_{n}^{\rm (LB)}$ of the rearranged
perturbation expansion (\ref{DLBmpt}),
i.e., the complete LB part of the expansion in logarithmic derivatives
(\ref{Dmpt}).
These coefficients in general differ from the coefficients
$d_n^{\rm (LB)}$ of the perturbation expansion (\ref{DLBpt})
in powers of $a(Q^2)$ which in general contain also some contributions
beyond large-$\beta_0$ (only at one-loop level
$d_{n}^{\rm (LB)}={\td}_n^{\rm (LB)}$). The result (\ref{DLBint})
is exactly renormalization scale independent; the scheme
dependence (i.e., dependence on the scheme coefficients
$c_j = \beta_j/\beta_0$, $j=2,\ldots,N-1$) appears, though, if
the coupling $a(t Q^2 e^{\cal C})$ there runs according to the
$N$-loop RGE ($N \geq 3$). Nonetheless, we will consider the
quantity (\ref{DLBint}) as a useful (quasi)observable and will
use it to test various evaluations of this quantity.
These evaluations will be based on the (artificially assumed) 
knowledge of only a finite number of terms in the expansion (\ref{DLBmpt}), 
i.e., resummations of truncated series where the couplings $a$ ($\A$)
and  $\ta_{n+1}$ ($\tA_{n+1}$) in these series are taken to be general
($N$-loop) couplings. 
The fact that all the terms of that series are known allows us
to evaluate the ``exact'' value of this quasiobservable and compare
it with the results of resummations of the truncated series.
This will give us indications of the quality of various resummation
methods, especially in frameworks with IR fixed point where we replace
$a \mapsto \A$ and $\ta_{n+1} \mapsto \tA_{n+1}$ in the above expressions
Eqs.~(\ref{DLB}). 
Since the resummation
methods are based on given {\it truncated\/} series, they in general
do not reproduce the correct large-$n$ behavior as dictated by the
renormalon structure. We are interested in the numerical efficiency of such
methods, i.e., the renormalization-scale (in)dependence and
the convergence behavior of the resummed results.

We mention here that there exist various other models for the
Adler function coefficients, such as the one used in 
Ref.~\cite{BMS3} which captures main features of the renormalon
growth and reproduces the full first three coefficients ($d_1, d_2, d_3$),
and renormalon models of Refs.~\cite{BeJa}. Furthermore, an approach
which allows generalization of the expression (\ref{DLB}) beyond the
large-$\beta_0$ approximation can be found in Ref.~\cite{Mikh}.
In this work, we chose the large-$\beta_0$ expression (\ref{DLB})
as the test case because of the practical simplicity of the
evaluation of the ``exact'' values, i.e., of the integral
(\ref{DLBint}).

The coefficients ${\td}_n^{\rm (LB)}$ can be represented as
logarithmic moments of the distribution function
$F_d(t)$ of the Adler function
\be
{\td}_n^{\rm (LB)} = (- \beta_0)^n \int_{t=0}^{\infty} d (\ln t)
\ln^n \left( t e^{{\cal C}} \right) F_d(t) \ .
\label{tdns}
\ee
For simplicity, we perform the
evaluations in the $c_2=c_3=\ldots = 0$ renormalization scheme,
where the pQCD running coupling $a(\kappa Q^2)$
has formally the two-loop form and is
expressed with the Lambert function  $W(z)$,
cf.~Refs.~\cite{Gardi:1998qr,Magr}.
Only in the analytic QCD model with two deltas (2$\delta$anQCD), 
Ref.~\cite{2danQCD}, 
we will use for the renormalization scheme the preferred central 
Lambert scheme of the model (with $N_f=3$): $c_2 = -4.76$, $c_j = c_2^{j-1}/c_1^{j-2}$ ($j=3,4,\ldots$),
where the exact solution of the underlying pQCD coupling is also known
in terms of the Lambert function (Ref.~\cite{Gardi:1998qr}, cf.~also
Ref.~\cite{CveKon}). We refer for some more details on this to Appendix
\ref{appLamb}.

\begin{figure}[htb] %\unitlength=1mm
\begin{minipage}[b]{.49\linewidth}
\centering\includegraphics[width=80mm]{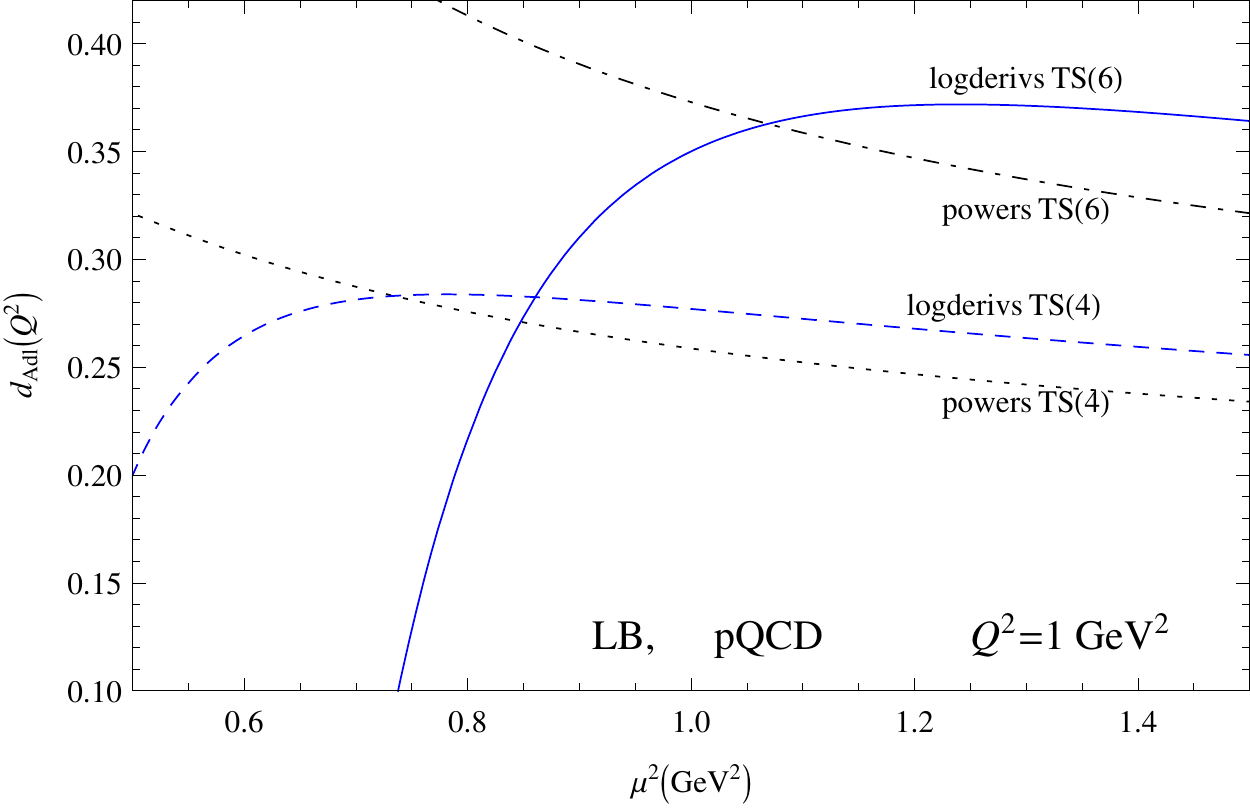}
%\centering\includegraphics[width=80mm]{LBpQCDvsmu2Q21.pdf}
%\centering{\epsfig{file=LBpQCDvsmu2Q21.eps,width=120mm,angle=0}}
\end{minipage}
\begin{minipage}[b]{.49\linewidth}
\centering\includegraphics[width=80mm]{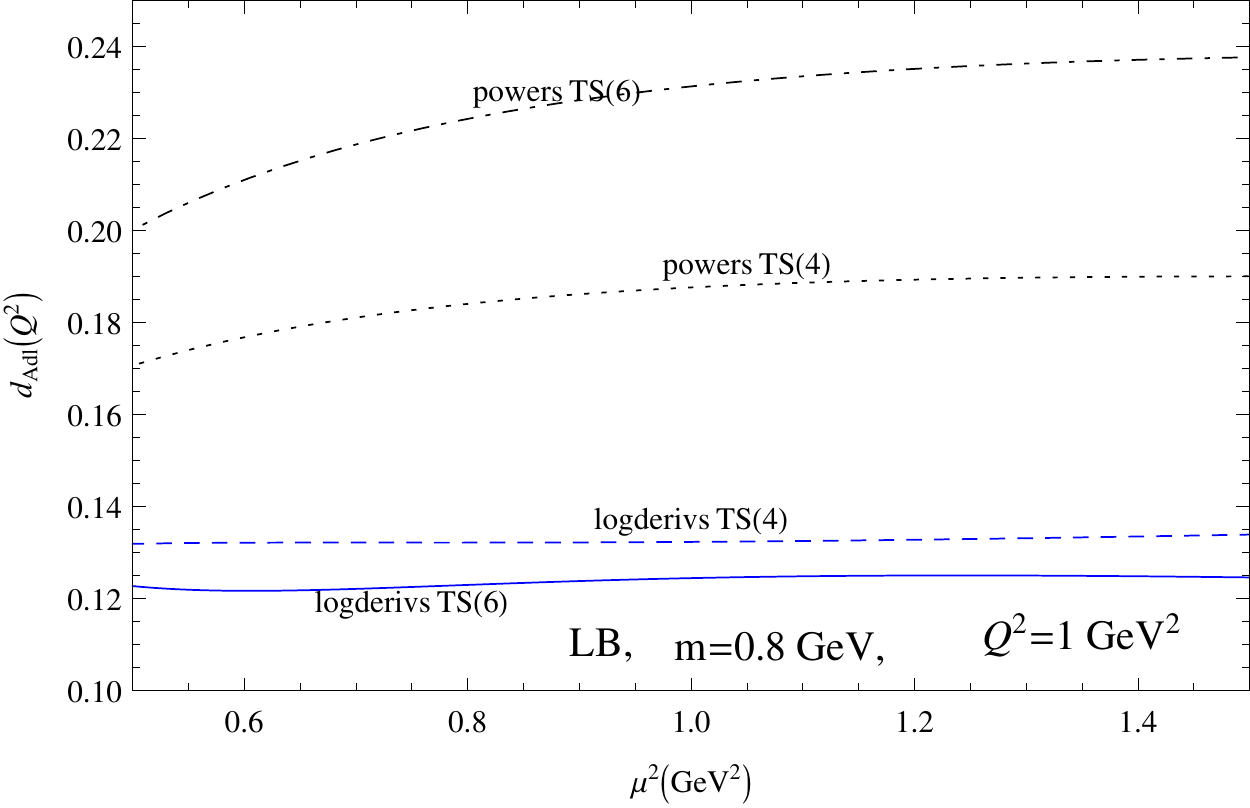}%
%\centering\includegraphics[width=80mm]{LBrhovsmu2Q21.pdf}
%\centering{\epsfig{file=LBrhovsmu2Q21.eps,width=80mm,angle=0}}
\end{minipage}
\begin{minipage}[b]{.49\linewidth}
\centering\includegraphics[width=80mm]{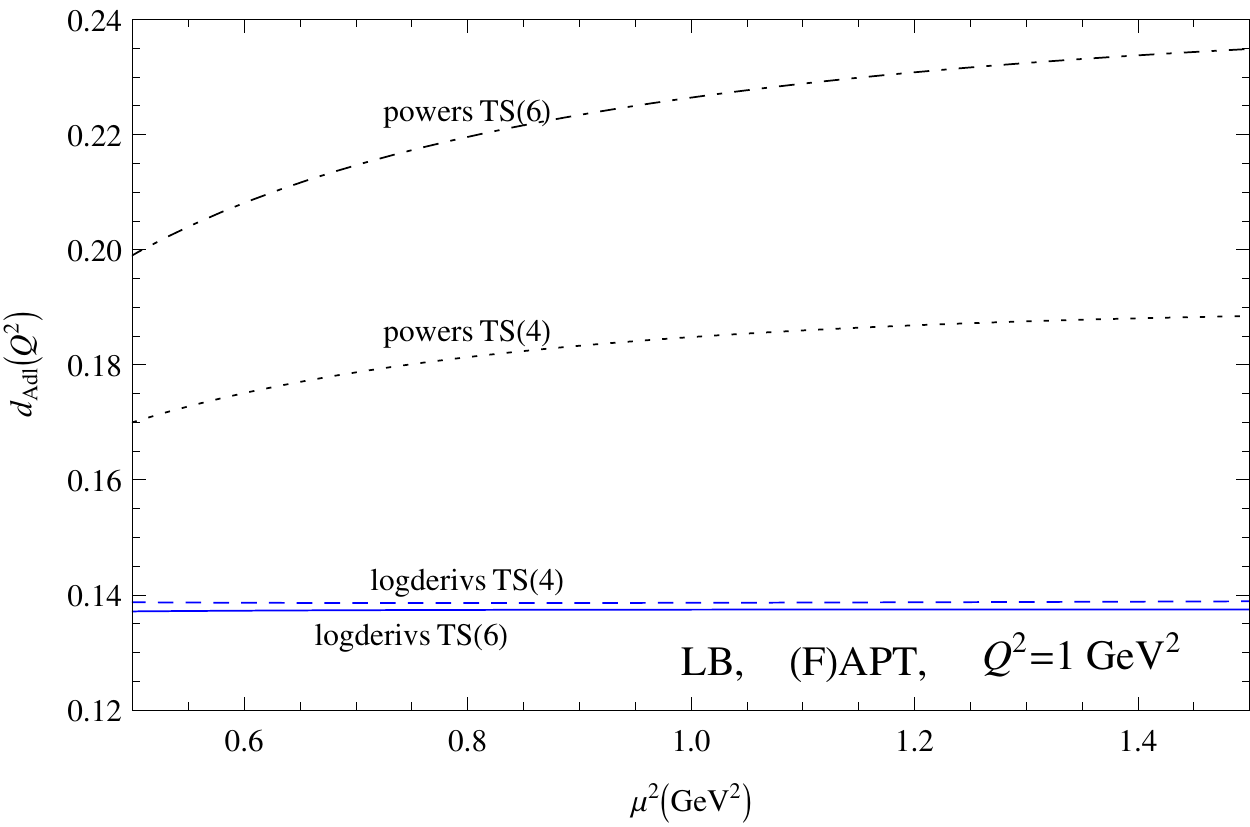}
%\centering\includegraphics[width=80mm]{LBMAvsmu2Q21.pdf}
%\centering{\epsfig{file=LBMAvsmu2Q21.eps,width=80mm,angle=0}}
\end{minipage}
\begin{minipage}[b]{.49\linewidth}
\centering\includegraphics[width=80mm]{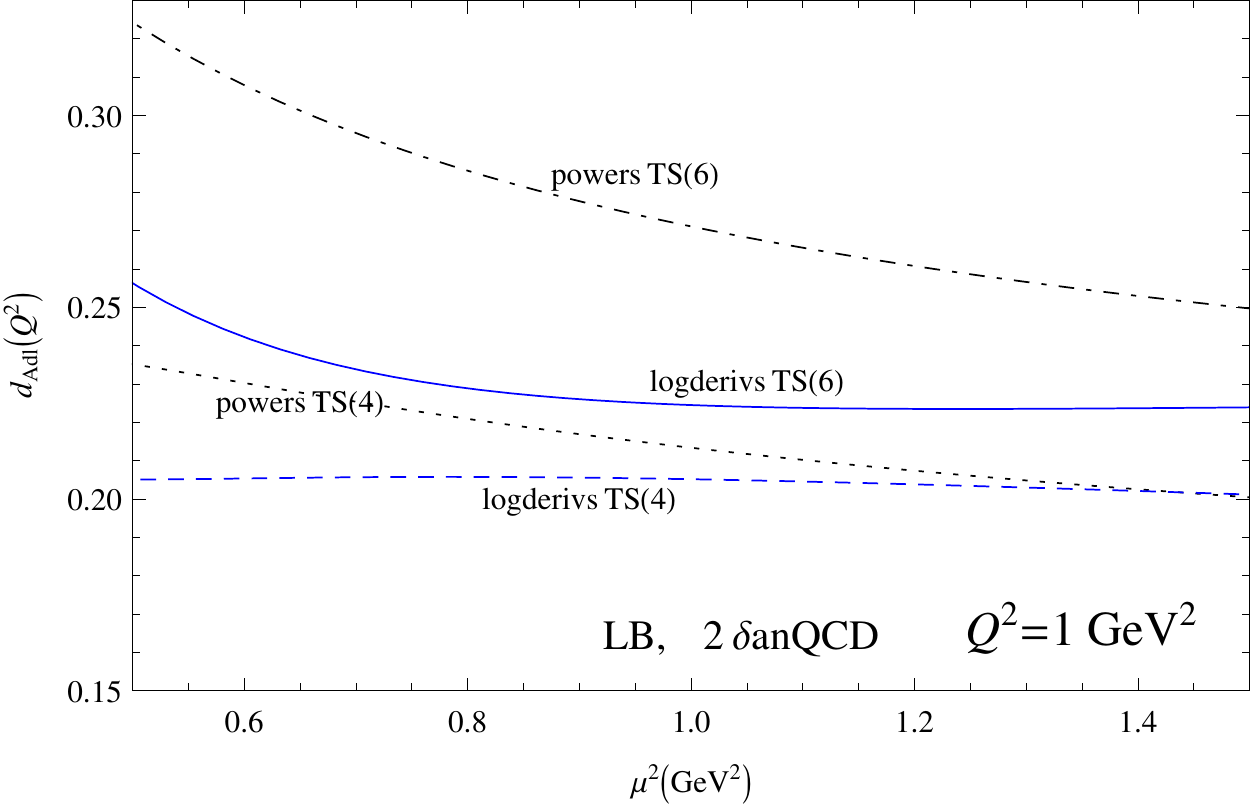}
%\centering\includegraphics[width=80mm]{LB2danvsmu2Q21.pdf}
%\centering\includegraphics[width=80mm]{LBanvsmu2Q21.pdf}
%centering{\epsfig{file=LBanvsmu2Q21.eps,width=80mm,angle=0}}
\end{minipage}
\vspace{-0.2cm}
 \caption{The effective charge of the massless 
Adler function $d_{\rm Adl}(Q^2)$,
at leading-$\beta_0$ (LB), for $Q^2=1 \ {\rm GeV}^2$, 
as a function of the (squared) spacelike renormalization scale $\mu^2$:
(a) in pQCD [Eq.~(\ref{aptexact})] (the upper left-hand Figure);
and in IR fixed point frameworks: 
(b) with the constant effective gluon mass 
$m=0.8$ GeV (the upper right-hand Figure); 
(c) the (fractional) analytic
perturbation theory (F)APT (the lower left-hand Figure); and (d) 
the analytic model 2$\delta$anQCD which has, in the discontinuity function
of $\A(Q^2)$, two delta functions in the low-$\sigma$ regime 
(the lower right-hand Figure). The truncations
are made at $\sim \A^4$ ($\tA_4$) and $\sim \A^6$ ($\tA_6$).}
\label{LBvsmu}
 \end{figure}
We vary the renormalization scheme $\mu^2 = \kappa Q^2$,
and perform evaluations in various IR fixed point
frameworks, with $N_f=3$: 
\begin{enumerate}
\item
A representative case of freezing -- the
case with constant effective gluon mass $m$, 
Refs.~\cite{Simonov,BKS,KKSh}, Eq.~(\ref{DS2l}),
applied to the coupling (\ref{aptexact}) in
the spirit of Ref.~\cite{Luna}
\be
\A^{(m)}(\mu^2) = a(\mu^2+m^2) \ ,
\label{rhocase}
\ee
where we take $m = 0.8$ GeV and $\Lambda_{\rm L.}=0.487$ GeV, giving at $\mu^2=m_{\tau}^2$ 
the value $\A(m_{\tau}^2) = 0.293/\pi$.
\item
The (fractional) analytic perturbation theory (F)APT case,
Refs.~\cite{ShS,MS,Sh1Sh2,BMS1,BMS2,BMS3}, 
Eq.~(\ref{MAAnudisp}). The APT scale is
fixed at ${\Lambda}_{\rm L.}({\rm APT}) = 0.572$ GeV, and $N_f=3$, giving the value
$\A_1^{\rm (APT)}(m_{\tau}^2)=0.295/\pi$.
\item
The analytic QCD case with two deltas (2$\delta$anQCD)
in the low-$\sigma$ region
for the analyticity function, Ref.~\cite{2danQCD},
Eqs.~(\ref{2dA1}) and (\ref{dif2d}). This model is 
numerically very close to pQCD coupling (\ref{aptLambexact}),
with the exception of the regime 
$|\mu^2| < 1 \ {\rm GeV}^2$. The input values
of the model are the central ones used in Ref.~\cite{2danQCD}
(among them: $c_2=-4.76$, $\Lambda_{\rm L.} = 0.260$ GeV)
and give the value $\A_1^{\rm (2 \delta)}(m_{\tau}^2)=0.291/\pi$.
\end{enumerate}
Furthermore, the first three full (i.e., LB+beyondLB)
coefficients $d_1$, $d_2$ and $d_3$ ($\Rightarrow$ $\td_1$, $\td_2$, $\td_3$)
of the Adler function are
now exactly known \cite{d1,d2,d3}
\bes
\label{dtpt}
\bea
d_{\rm Adl}(Q^2)_{\rm pt}^{[4]}
&=& a(Q^2) + d_1 a(Q^2)^2 + d_2 a(Q^2)^3 + d_3 a(Q^2)^4 \ ,
\\
d_{\rm Adl}(Q^2)_{\rm mpt}^{[4]}
&=& a(Q^2) + \td_1 \ta_2(Q^2) + \td_2 \ta_3(Q^2) + \td_3 \ta_4(Q^2) \ .
\eea
\ees
That is why we can also evaluate the full Adler function,
but truncated at order 4 (TS[4]) or lower, in any
scheme and at any scale $\mu^2=\kappa Q^2$, for example
in pQCD and in the aforementioned four IR fixed point frameworks.

\begin{figure}[htb] %\unitlength=1mm
\begin{minipage}[b]{.49\linewidth}
\centering\includegraphics[width=80mm]{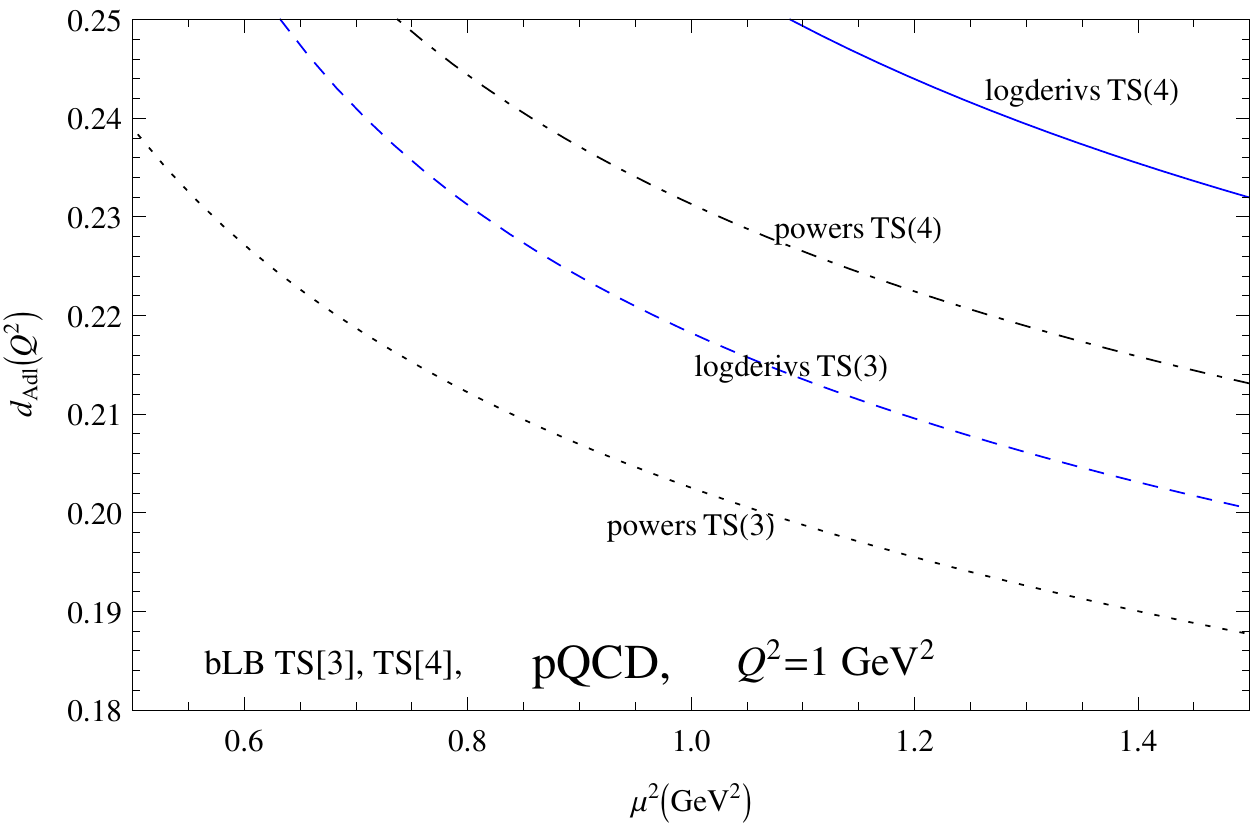}
%\centering\includegraphics[width=80mm]{pQCDvsmu2Q21.pdf}
%\centering{\epsfig{file=pQCDvsmu2Q21.eps,width=120mm,angle=0}}
\end{minipage}
\begin{minipage}[b]{.49\linewidth}
\centering\includegraphics[width=80mm]{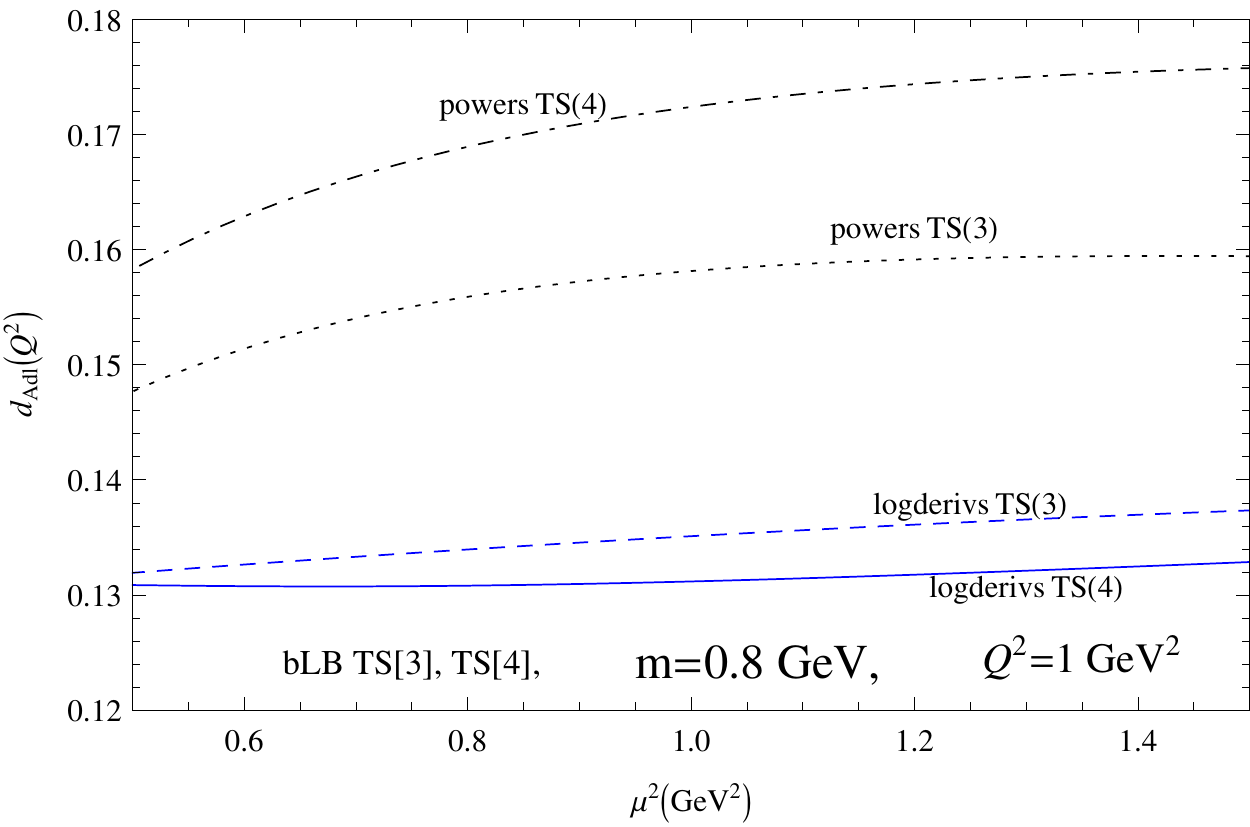}
%\centering\includegraphics[width=80mm]{rhovsmu2Q21.pdf}
%centering{\epsfig{file=rhovsmu2Q21.eps,width=80mm,angle=0}}
\end{minipage}
\begin{minipage}[b]{.49\linewidth}
\centering\includegraphics[width=80mm]{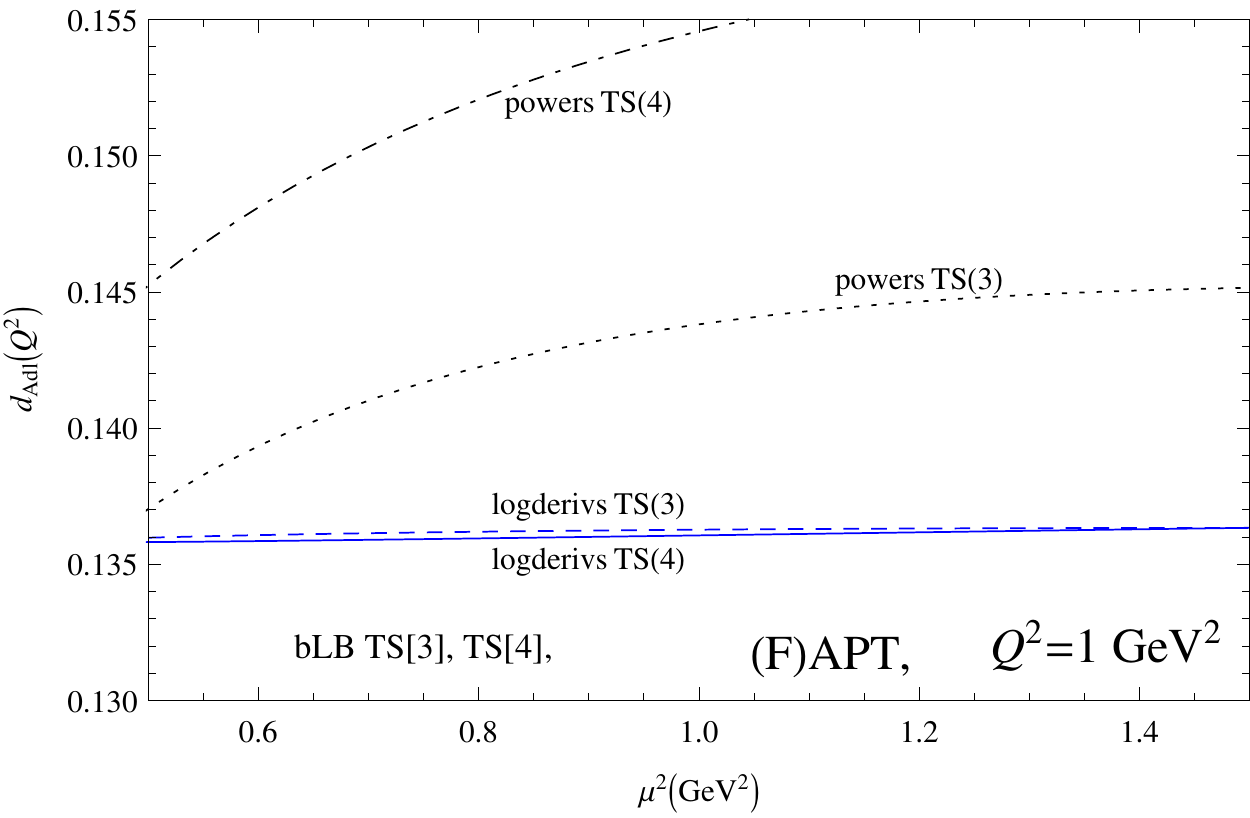}
%\centering\includegraphics[width=80mm]{MAvsmu2Q21.pdf}
%centering{\epsfig{file=MAvsmu2Q21.eps,width=80mm,angle=0}}
\end{minipage}
\begin{minipage}[b]{.49\linewidth}
\centering\includegraphics[width=80mm]{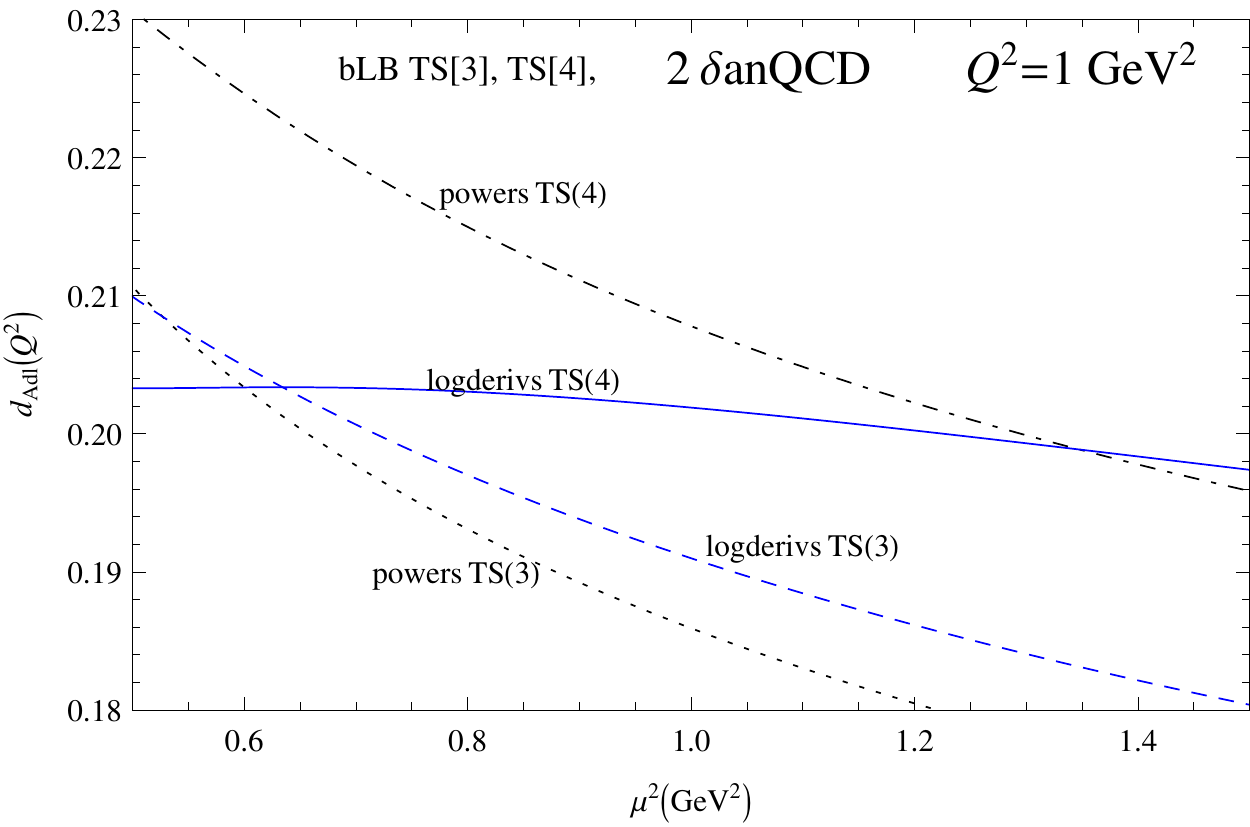}
%\centering\includegraphics[width=80mm]{2danvsmu2Q21.pdf}
\end{minipage}\vspace{-0.2cm}
 \caption{The same as in Fig.~\ref{LBvsmu}, but for the
truncated series with the full (LB+beyondLB) coefficients,
cf.~Eqs.~(\ref{dtpt}). The truncations are made at
 $\sim \A^3$ ($\tA_3$) and $\sim \A^4$ ($\tA_4$).}
\label{vsmu}
 \end{figure}
The results of the LB Adler function, truncated at
order 4 and 6, as power series and as series in logarithmic derivatives,
for $Q^2 = 1 \ {\rm GeV}^2$, are presented
as functions of the squared (spacelike) renormalizations scale 
$\mu^2 = \kappa Q^2$ in Figs.~\ref{LBvsmu} for
the pQCD case and
for the three considered aforementioned IR fixed point frameworks.
Truncations are made at $\sim \A^4$ and  $\sim \A^6$ for power
series, and at $\tA_4$ and $\tA_6$ for the series in
logarithmic derivatives.

Furthermore, the analogous results based on the truncated series 
(\ref{dtpt}) with full (LB+bLB) coefficients, are given 
in Figs.~\ref{vsmu} for
pQCD and for the three considered
IR fixed point cases.
Truncations are made at $\sim \A^3$ ($\tA_3$) and  $\sim \A^4$ ($\tA_4$).

These figures show how
the arguments of the previous Section manifest
themselves in practice. 
In the IR fixed point frameworks, the
truncated power expansions have increasingly strong
renormalization scale dependence, due to the
wrong incorporation of the nonperturbative contributions
at higher orders there. This effect is stronger
when $Q^2$ values are lower. 
On the other hand, the truncated series
in logarithmic derivatives, in the IR fixed point frameworks,
have weaker scale dependence,
and this dependence in general does not get stronger when
the number of terms in the truncated series increases.
Furthermore, these figures indicate
that the power series has 
divergent behavior already at relatively low orders, in contrast
to the series in logarithmic derivatives.

On the other hand,
in pure pQCD scenario, the two types of truncated series
give comparable results and it is not clear which one
is better, as demonstrated also in Ref.~\cite{GCCVMar}.

\subsection{Convergence properties}
\label{subs:conv}

In this Subsection we present, for the leading-$\beta_0$ (LB)
Adler function as a test case, the convergence properties
of truncated series in powers, in logarithmic derivatives,
and of a resummed version of the latter series based on a
generalized diagonal Pad\'e (dPA) method. The latter method
\begin{figure}[htb] %\unitlength=1mm
\begin{minipage}[b]{.49\linewidth}
\centering\includegraphics[width=80mm]{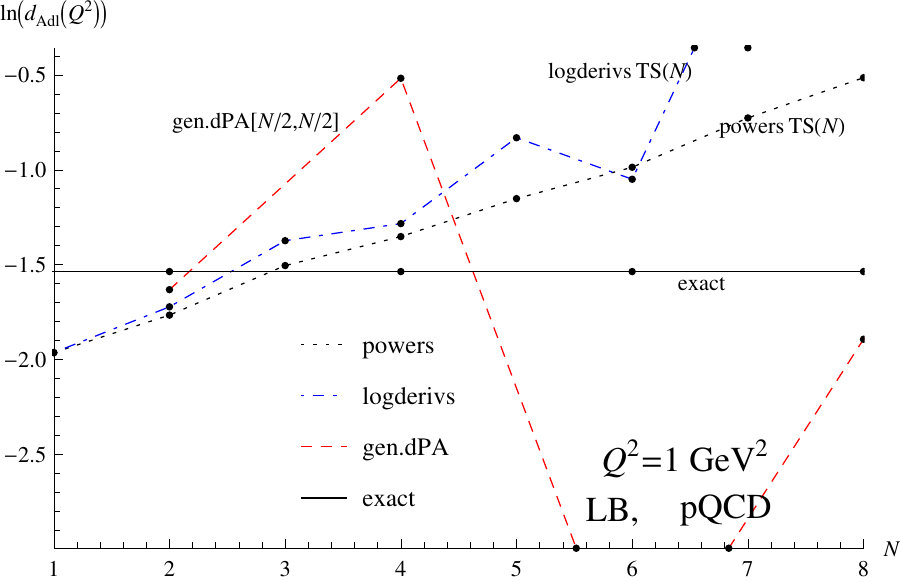}
%\centering\includegraphics[width=80mm]{figdvLBdBGptA.pdf}
%centering{\epsfig{file=figdvLBdBGptA.eps,width=80mm,angle=0}}
\end{minipage}
\begin{minipage}[b]{.49\linewidth}
\centering\includegraphics[width=80mm]{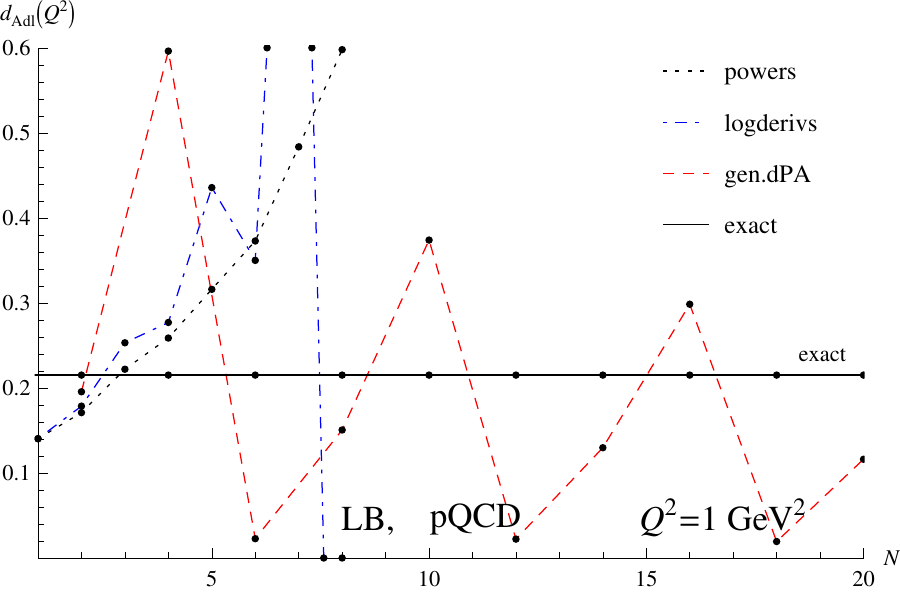}
%\centering\includegraphics[width=80mm]{figdvLBdBGptB.pdf}
%centering{\epsfig{file=figdvLBdBGptB.eps,width=80mm,angle=0}}
\end{minipage}
\vspace{-0.2cm}
 \caption{The convergence (divergence) behavior of the LB Adler function
at $Q^2 = 1 \ {\rm GeV}^2$, as a function of the truncation order $N$, 
in pQCD. The left-hand Figure is for a larger interval of values of 
the LB Adler function - the vertical axis represents 
$\ln d_{\rm Adl}^{\rm (LB)}(Q^2)$. The right-hand Figure is for a
narrower interval of values $ d_{\rm Adl}^{\rm (LB)}(Q^2)$, and for a 
larger $N$-interval. See the text for other details.}
\label{dvLBdBGpt}
 \end{figure}
\begin{figure}[htb] %\unitlength=1mm
\begin{minipage}[b]{.49\linewidth}
\centering\includegraphics[width=80mm]{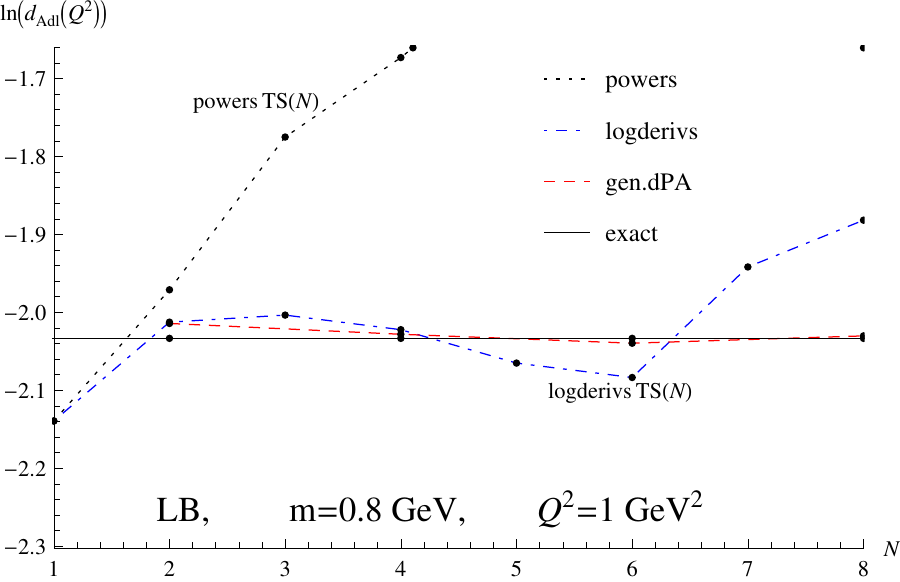}
%\centering\includegraphics[width=80mm]{figdvLBdBGrhoA.pdf}
%centering{\epsfig{file=figdvLBdBGrhoA.eps,width=80mm,angle=0}}
\end{minipage}
\begin{minipage}[b]{.49\linewidth}
\centering\includegraphics[width=80mm]{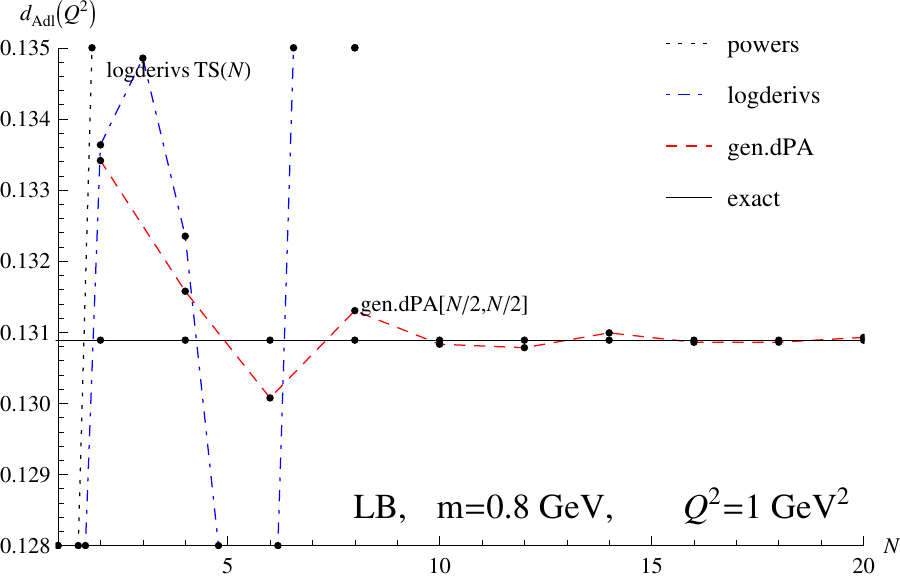}
%\centering\includegraphics[width=80mm]{figdvLBdBGrhoB.pdf}
%centering{\epsfig{file=figdvLBdBGrhoB.eps,width=80mm,angle=0}}
\end{minipage}
\vspace{-0.2cm}
 \caption{The same as in Fig.~\ref{dvLBdBGpt}, but for the
case of IR fixed point scenario with the constant gluon effective mass
$m=0.8$ GeV.}
\label{dvLBdBGm}
 \end{figure}
was introduced in Ref.~\cite{BGApQCD1} in the context of
pQCD, and was motivated by the dPA resummation approach and
its renormalization scale independence in the one-loop approximation
\cite{Gardi}.
It was later applied to analytic QCD frameworks
in Refs.~\cite{GCRK} and \cite{anOPE}. It consists in the
following expression:
\be
{\cal G}^{[M/M]}_{d}(Q^2) =  
\sum_{j=1}^M \tal_j \; \A(\kappa_j Q^2) \ ,
\label{dBG}
\ee
where the scale parameters $\kappa_j$ and
the coefficients $\tal_j$ ($\tal_1 + \ldots + \tal_M = 1$)
are determined uniquely from the known truncated series
of the observable $d(Q^2)$ up to ${\ta}_{2 M}$
($\sim a^{2 M}$)
\be
d(Q^2; \mu^2)_{\rm mpt}^{[2 M]} = a(\mu^2) + 
\sum_{j=1}^{2 M - 1} {\td}_j(\mu^2/Q^2) \; {\ta}_{j+1}(\mu^2) \ .
\label{Dtmpt}
\ee
The mentioned parameters $\kappa_j$ and $\tal_j$ are obtained
by regarding the series (\ref{Dtmpt}) in logarithmic derivatives
as formally a (truncated) series in powers of one-loop coupling 
[${\ta}_{j+1}(\mu^2) \mapsto a_{1\ell}(\mu^2)^{j+1}$]
\be
{\widetilde d}(Q^2; \mu^2)_{\rm pt}^{[2 M]} = a_{1\ell}(\mu^2) + 
\sum_{j=1}^{2 M -1} {\td}_j(\mu^2/Q^2) \; a_{1\ell}(\mu^2)^{j+1} \ ,
\label{tDpt}
\ee 
and constructing for it the diagonal Pad\'e (dPA) $[M/M]$ which\footnote{
$[M/M]_{\widetilde d}$ is by definition a ratio of two polynomials in 
$a _{1\ell}(\mu^2)$ of order $M$ each,
and whose coefficients are determined by the condition: 
$[M/M]_{\widetilde d} - {\widetilde d}(Q^2;\mu^2)_{\rm pt}^{[2 M]} \sim a _{1\ell}^{2 M+1}$.}
 is then
decomposed in a linear combination of simple fractions
%\footnote{In Mathematica software \cite{Math}, the command ``Apart'' achieves this decomposition.}
(in Mathematica software \cite{Math}, 
the command ``Apart'' achieves this decomposition)
\be
[M/M]_{\widetilde d}(a_{1\ell}(\mu^2)) =  
\sum_{j=1}^M \tal_j \frac{x}{1 + \tu_j x}{\bigg |}_{x=a_{1\ell}(\mu^2)} 
\ .
\label{MMdecom1}
\ee
Each simple fraction $x/(1+ \tu_j x)$ [with: $x=a_{1\ell}(\mu^2)$]
can be written as $a_{1 \ell}(\kappa_j Q^2)$, i.e.,
\be
[M/M]_{\widetilde d}(a_{1\ell}(\mu^2)) = 
  \sum_{j=1}^M \tal_j \; a_{1\ell}(\kappa_j Q^2) \ , \quad 
{\rm where \ } \kappa_j Q^2 = \mu^2 \exp(\tu_j/\beta_0) \ .
\label{MMdecom2}
\ee
This procedure gives us the mentioned parameters $\tal_j$ and $\kappa_j$;
it turns out that they are exactly-independent of the chosen
renormalization scale $\mu^2$, Refs.~\cite{BGApQCD1,anOPE},
and that the resummed conformal
approximant ${\cal G}^{[M/M]}_{d}(Q^2)$, Eq.~(\ref{dBG}),
fulfills the basic requirement of the 
approximant of order $N=2 M$, Ref.~\cite{GCRK}
\be
d(Q^2) - {\cal G}^{[M/M]}_{d}(Q^2) = {\cal O}(\tA_{2 M+1})
= {\cal O}(\A_{2 M+1})
\ .
\label{dBGappr}
\ee
We stress that the approximant (\ref{dBG}) is applicable in the general case
of $N$-loop RGE-running of $\A(\mu^2)$, and the relation (\ref{dBGappr})
is valid in such general case.\footnote{
It is valid for any RGE running with a given beta function, and this
function can contain terms nonanalytic in $\A$, as it happens in
the models with IR fixed point.}
This is what makes this approximant so
attractive theoretically, as pointed out in Refs.~\cite{BGApQCD1,GCRK,anOPE}.
It is thus not the direct dPA method of Ref.~\cite{Gardi}, 
but a nontrivial generalization
thereof, which takes into account the general $N$-loop RGE-running of the
couplings and gives exactly renormalization scale independent results.
The crucial part in the construction of this method is the
formal replacement $\ta_{j+1} \mapsto a^{j+1}$ [Eqs.~(\ref{Dtmpt})-(\ref{tDpt})]
as an intermediate step, and the use of dPA on the formal
power series to obtain the scale and weight parameters
$\kappa_j$ and $\tal_j$.

As shown in Refs.~\cite{GCRK,anOPE}, these approximants work very
well in practice in the analytic QCD frameworks.

\begin{figure}[htb] %\unitlength=1mm
\begin{minipage}[b]{.49\linewidth}
\centering\includegraphics[width=80mm]{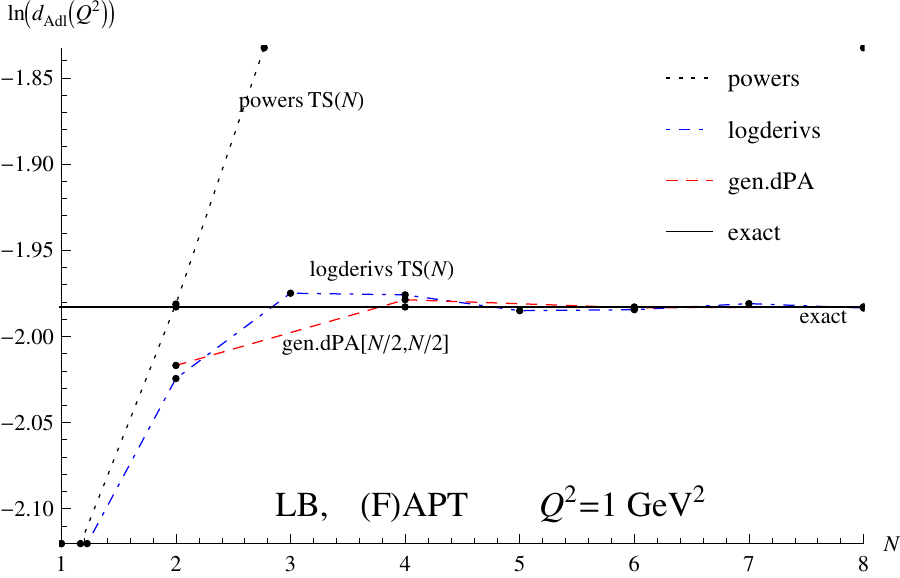}
%\centering\includegraphics[width=80mm]{figdvLBdBGMAA.pdf}
%centering{\epsfig{file=figdvLBdBGMAA.eps,width=80mm,angle=0}}
\end{minipage}
\begin{minipage}[b]{.49\linewidth}
\centering\includegraphics[width=80mm]{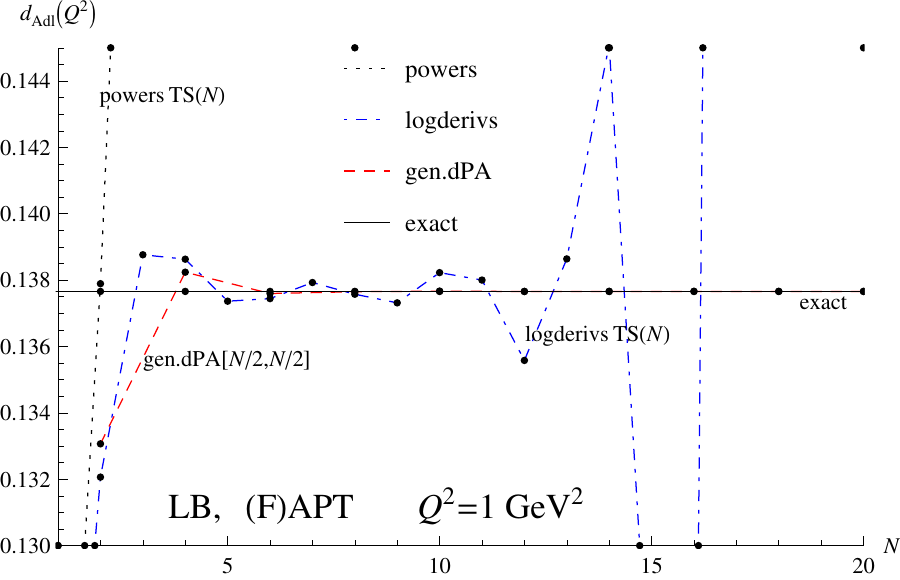}
%\centering\includegraphics[width=80mm]{figdvLBdBGMAB.pdf}
%centering{\epsfig{file=figdvLBdBGMAB.eps,width=80mm,angle=0}}
\end{minipage}
%\vspace{-0.2cm}
% \caption{The same as in Fig.~\ref{dvLBdBGpt}, but for the
%case of APT model.}
%\label{dvLBdBGAPT}
% \end{figure}
%\begin{figure}[htb] %\unitlength=1mm
\begin{minipage}[b]{.49\linewidth}
\centering\includegraphics[width=80mm]{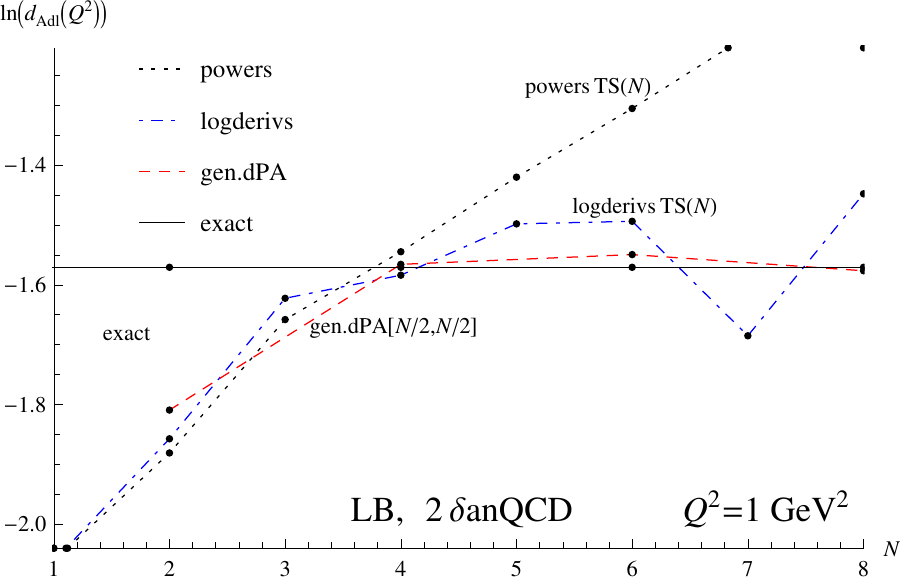}
%\centering\includegraphics[width=80mm]{figdvLBdBG2danA.pdf}
%\centering\includegraphics[width=80mm]{figdvLBdBGanA.pdf}
%centering{\epsfig{file=figdvLBdBGanA.eps,width=80mm,angle=0}}
\end{minipage}
\begin{minipage}[b]{.49\linewidth}
\centering\includegraphics[width=80mm]{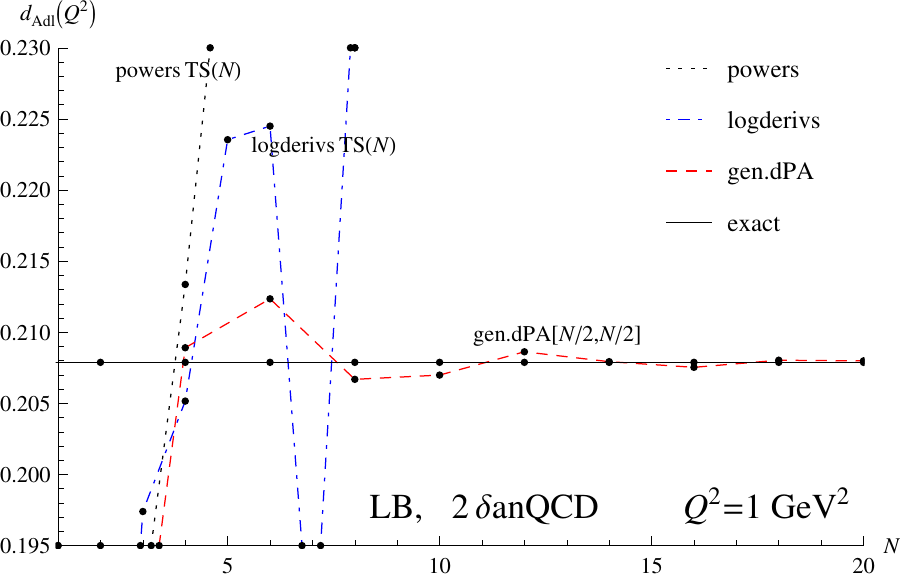}
%\centering\includegraphics[width=80mm]{figdvLBdBG2danB.pdf}
%\centering\includegraphics[width=80mm]{figdvLBdBGanB.pdf}
%centering{\epsfig{file=figdvLBdBGanB.eps,width=80mm,angle=0}}
\end{minipage}
\vspace{-0.2cm}
 \caption{The same as in Fig.~\ref{dvLBdBGpt}, but for the
case of the APT model (the upper two Figures) and
the 2$\delta$ analytic QCD model (the lower two Figures).}
%\label{dvLBdBGan}
\label{dvLBdBGAPTan}
 \end{figure}
Using the LB Adler function, 
Eqs.~(\ref{DLB}) and (\ref{tdns}), as a test case,\footnote{
The authors of Refs.~\cite{ananth} evaluated the Adler function and
related quantities in pQCD framework using nonpower expansions
based on conformal transformations and renormalon structure of the
Borel transform, and using as test quantities renormalon models for 
Adler function of Refs.~\cite{BeJa}.}
at $Q^2 = 1 \ {\rm GeV}^2$,
we present in Figs.~\ref{dvLBdBGpt} the results of the
evaluation of this ``quasi-observable'' as a function of 
the truncation order $N$ in the case of pQCD coupling $a$, Eq.~(\ref{aptexact}),
as: truncated power series, truncated series in logarithmic
derivatives, and the generalized dPA Eq.~(\ref{dBG})
(in that case: $N=2 M = 2, 4,\ldots$). We can see that the power
series and the series in logarithmic derivatives increase with
increasing $N$ above the exact value\footnote{The ``exact'' value is here
taken as the Principal Value of the integral (\ref{DLBint}) which
has ambiguity due to Landau singularities of pQCD coupling. No such
ambiguity problems appear in the other considered cases, because they have IR fixed point.},
while the generalized dPA oscillates uncontrollably around it.

In Figs.~\ref{dvLBdBGm} we present the corresponding results
for the gluon effective mass case: $m=0.8$ GeV 
case of Eq.~(\ref{rhocase}); in Figs.~\ref{dvLBdBGAPTan}
the results for the APT model of Eq.~(\ref{MAAnudisp})
%; finally, in Figs.~\ref{dvLBdBGan}
%we present the results for the 2$\delta$anQCD model of
%Eqs.~(\ref{2dA1}) and (\ref{dif2d}).
and of the 2$\delta$anQCD model of Eqs.~(\ref{2dA1}) and (\ref{dif2d}).

We can see that, in contrast to pQCD, any framework with
IR fixed point gives for the series in logarithmic derivatives
a clearly better convergence properties than for the power series. 
The power series, although having usually $\A(Q^2) < 1$, 
is badly divergent, in part due to a 
renormalon growth of the coefficients
$d_n$ ($\sim n!$). However, we note that the series in 
logarithmic derivatives also has a
(one-loop) renormalon growth of the coefficients $\td_{n}$. 
And both the power terms and the logarithmic derivatives, at any $Q^2$, have the
hierarchy: $\A(Q^2) > \A(Q^2)^2 > \A(Q^2)^3 > \ldots$ and 
$\A(Q^2) >  |\tA_2(Q^2)| >  |\tA_3(Q^2)| > \ldots$. Nonetheless, the 
logarithmic derivatives $\tA_n(Q^2)$ have alternating signs at large $n$,
which indicates why the series in logarithmic derivatives has a better
convergence (less severe divergence) behavior than the power series.

The results of the Figures further indicate 
that the generalized dPA method works very well in all 
the frameworks with IR fixed point, it gives a clearly
convergent behavior when the truncation order $N$ ($=2 M$) increases. 

\begin{figure}[htb] %\unitlength=1mm
\begin{minipage}[b]{.49\linewidth}
\centering\includegraphics[width=80mm]{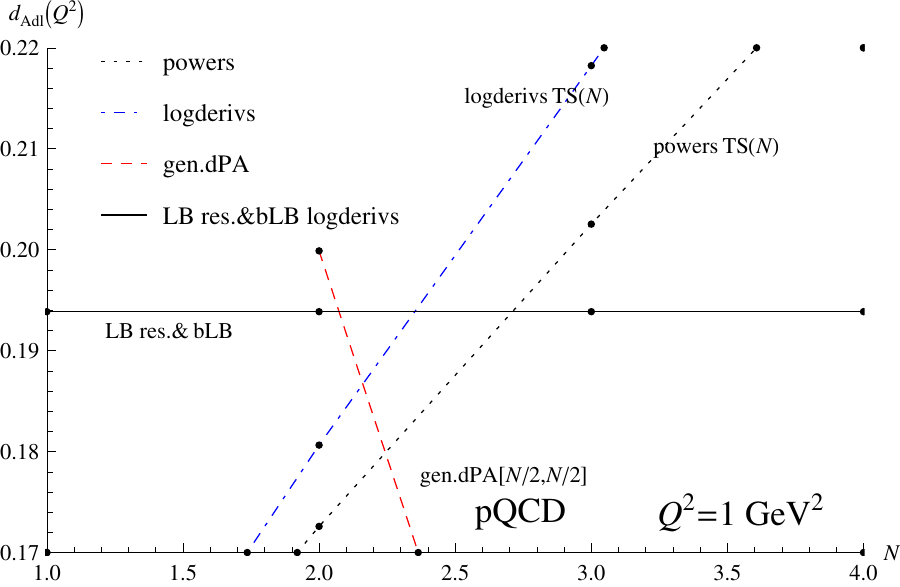}
%\centering\includegraphics[width=80mm]{figdvpt.pdf}
%centering{\epsfig{file=figdvpt.eps,width=80mm,angle=0}}
\end{minipage}
\begin{minipage}[b]{.49\linewidth}
\centering\includegraphics[width=80mm]{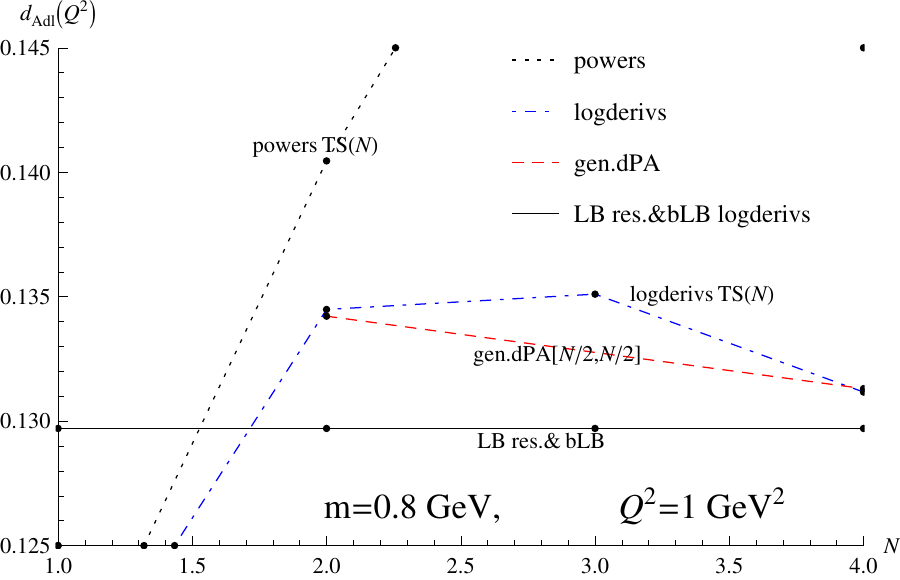}
%\centering\includegraphics[width=80mm]{figdvrho.pdf}
%centering{\epsfig{file=figdvrho.eps,width=80mm,angle=0}}
\end{minipage}
\begin{minipage}[b]{.49\linewidth}
\centering\includegraphics[width=80mm]{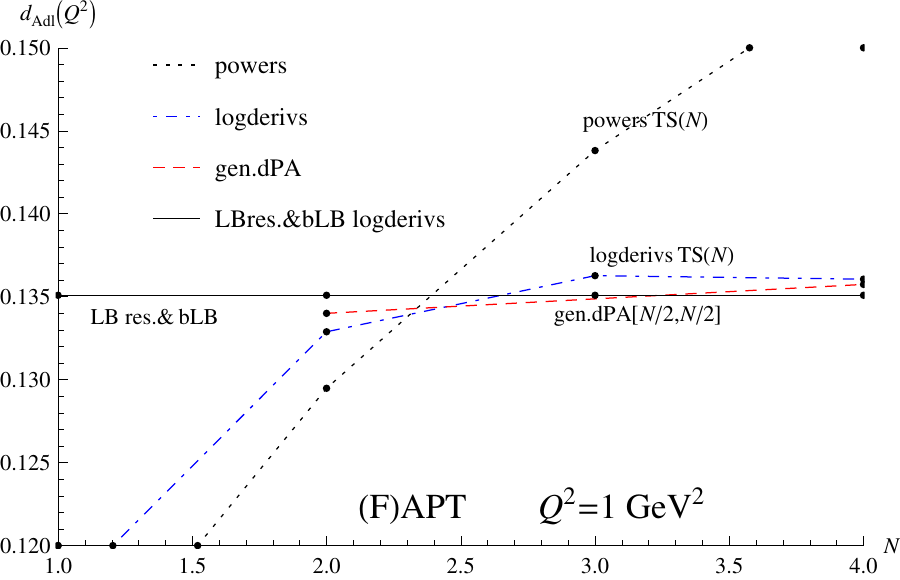}
%\centering\includegraphics[width=80mm]{figdvMA.pdf}
%centering{\epsfig{file=figdvMA.eps,width=80mm,angle=0}}
\end{minipage}
\begin{minipage}[b]{.49\linewidth}
\centering\includegraphics[width=80mm]{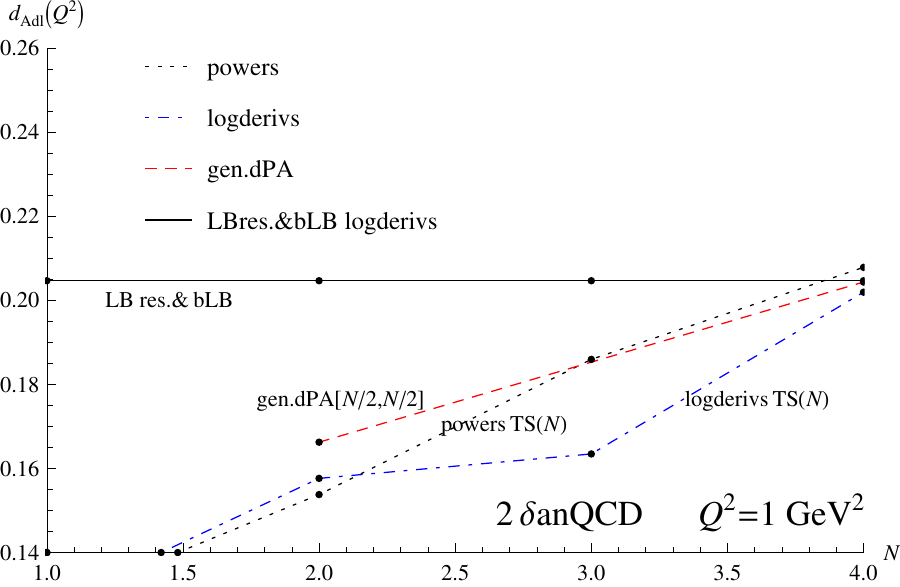}
%\centering\includegraphics[width=80mm]{figdv2dan.pdf}
%\centering\includegraphics[width=80mm]{figdvan.pdf}
%\centering{\epsfig{file=figdvan.eps,width=80mm,angle=0}}
\end{minipage}\vspace{-0.2cm}
 \caption{Analogous results as those of 
Figs.~\ref{dvLBdBGpt}-\ref{dvLBdBGAPTan}, but for the
truncated series based on the full (LB+beyondLB) coefficients,
cf.~Eqs.~(\ref{dtpt}): for pQCD and for the three 
considered IR fixed point frameworks.}
\label{dvdBG}
 \end{figure}
Finally, in Fig.~\ref{dvdBG} we present analogous results as in
Figs.~\ref{dvLBdBGpt}-\ref{dvLBdBGAPTan}, for pQCD and
the three considered IR fixed point scenarios,
but this time with the known complete
(LB+beyondLB) coefficients $d_n$ ($\td_n$), cf.~Eq.~(\ref{dtpt}).
Since only up to $d_4$ ($\td_4$) coefficients
are known exactly, the results are shown only up to the
order $N=4$. Also in this
(LB+beyondLB) case, we can see that in the IR fixed point frameworks
the series in logarithmic derivatives behave significantly better than the
corresponding power series; and that the generalized dPA method
is often even better. These Figures include also the result of the
LB resummation [i.e., the integral (\ref{DLBint})]\footnote{
In the case of pQCD, the LB-integral has ambiguity due to the
Landau singularities, and we took the Principal Value in this case.}
with the three known beyond-LB terms added (here added in the form of
logarithmic derivatives). The latter method is also considered
as probably competitive with the generalized dPA method, at
least at the considered order ($N=4$). 

On the other hand, in pQCD it appears to be impossible to identify
a method that is better than the other methods.

\section{Evaluation of timelike physical quantities in IR fixed
point scenarios}
\label{timelike}

The extension of the described formalism to the evaluation
of timelike physical quantities ${\cal T}(s)$ ($s=-Q^2 > 0$)
is based on the assumption of existence of an integral transformation 
which relates such timelike quantities with the (corresponding) 
spacelike quantities ${\cal F}(Q^2)$. The latter are
evaluated in the aforedescribed way, for any complex $Q^2$,
and the integral trasformation is applied on them. 

\begin{figure}[htb] %\unitlength=1mm
\centering\includegraphics[width=60mm]{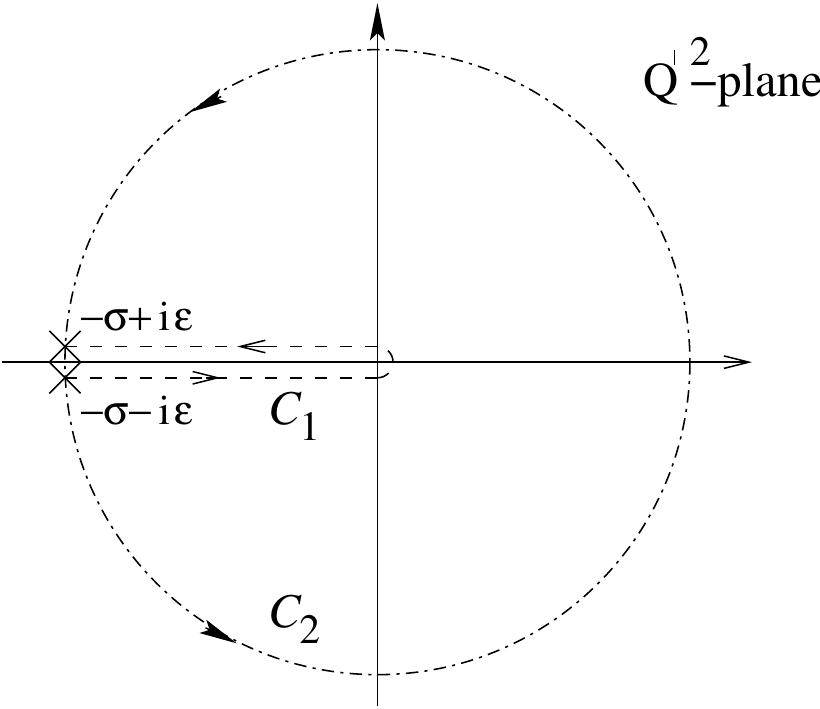}
%\centering\includegraphics[width=60mm]{contour12.jpg}
%\centering\includegraphics[width=60mm]{figrec2.pdf}
%\centering\epsfig{file=figrec2.eps,width=6.cm}
\vspace{-0.4cm}
 \caption{\footnotesize  Paths ${\cal C}_1$ and ${\cal C}_2$
in the complex $Q^{' 2}$-plane.}
\label{contour12}
 \end{figure}

Often the integral transformation is the same as
when ${\cal T}(s)$ is the $(e^+ e^- \to \ {\rm hadrons})$ ratio $R(s)$ and
${\cal F}(Q^2)$ is the Adler function (log-derivative of the
quark-current correlator)
\begin{equation}
{\cal F}(Q^2) = Q^2 \int_0^{\infty} 
\frac{d \sigma \ {\cal T}(\sigma)}{(\sigma + Q^2)^2} \ .
\label{FT}
\end{equation}
The inverse transformation is
\begin{equation}
{\cal T}(\sigma) = \frac{1}{2 \pi i} 
\int_{-\sigma - i \varepsilon}^{-\sigma + i \varepsilon} 
\frac{d Q^{' 2}}{Q^{' 2}} {\cal F}(Q^{' 2}) \ ,
\label{TF}
\end{equation}
where the integration contour is in the complex 
$Q^{' 2}$-plane encircling the singularities of the integrand,
e.g., path ${\cal C}_1$ or ${\cal C}_2$ of Fig.~\ref{contour12}.  
Let us consider the case when 
the truncated perturbation expansion of the (massless) spacelike quantity
${\cal F}(Q^2)$ in pQCD is of the form
\be
{\cal F}(Q^2;\kappa)^{[N]}_{\rm pt} = a(\kappa Q^2) + \sum_{n-1}^{N-1} {\cal F}_n(\kappa) a(\kappa Q^2)^{n+1} \ .
\label{calFpt}
\ee 
In IR fixed point scenarios this implies the following nonpower expansion,
as explained in the previous Section:
\be
{\cal F}(Q^2)^{[N]}_{\rm an} = {\A}(\kappa Q^2) + 
\sum_{n-1}^{N-1} {\widetilde {\cal F}}_n(\kappa) {\tA}_{n+1}(\kappa Q^2)
= {\A}(\kappa Q^2) + 
\sum_{n}^{N-1} {{\cal F}}_n(\kappa) {\A}_{n+1}(\kappa Q^2) \ .
\label{Fana}
\ee
The application of the integral transformation (\ref{TF}) to this
expression then gives the desired result 
\begin{equation}
{\cal T}(\sigma; \kappa)^{[N]}_{\rm an} = \frac{1}{2 \pi i} 
\int_{-\sigma - i \varepsilon}^{-\sigma + i \varepsilon} 
\frac{d Q^{' 2}}{Q^{' 2}} {\cal F}(Q^{' 2}; \kappa)^{[N]}_{\rm an} \ ,
\label{calTan1}
\end{equation}
This can be performed term-by-term, leading to
\be
{\cal T}(\sigma; \kappa)^{[N]}_{\rm an} = {\ntH}(\kappa \sigma) + 
\sum_{n-1}^{N-1} {{\cal F}}_n(\kappa) {\ntH}_{n+1}(\kappa \sigma) \ ,
\label{calTan2}
\ee 
where the timelike (Minkowskian) couplings ${\ntH}_{n}(\sigma)$ 
(with: ${\ntH} \equiv {\ntH}_1$ and $\A \equiv \A_1$) are defined as
\begin{equation}
{\ntH}_{n}(\sigma) \equiv \frac{1}{2 \pi i} 
\int_{-\sigma - i \varepsilon}^{-\sigma + i \varepsilon} 
\frac{d Q^{' 2}}{Q^{' 2}} {\A}_{n}(Q^{' 2}) \ ,
\label{ntHa}
\end{equation}
and the inverse transformation is
\be
\A_{n}(\kappa Q^2) = \kappa Q^2 \int_0^{\infty} 
\frac{d \sigma \; {\ntH}_{n}(\sigma)}{(\sigma + \kappa Q^2)^2} \quad
(n=1,2,\ldots) \ .
\label{AntHA}
\ee 
We can also apply the generalized dPA method mentioned in
Sec.~\ref{subs:conv} to evaluate
${\cal F}(Q^2; \kappa)^{[2 M]}$
\bes
\label{gendPAtl}
\bea
{\cal F}(Q^2; \kappa)^{[2 M]}& \mapsto & {\cal G}_{\cal F}(Q^2) =
\sum_{j=1}^M {\widetilde \alpha}_j \A(\kappa_j Q^2) \ ,
\\
\Rightarrow 
{\cal T}(\sigma; \kappa)^{[2 M]} & 
\mapsto & {\cal G}_{\cal T}(\sigma) = 
\sum_{j=1}^M {\widetilde \alpha}_j {\ntH}(\kappa_j \sigma) \ .
\eea
\ees

For example, to calculate the effective charge ${\cal T}(s) = r_{e^+ e^-}(s)$ of
the $(e^+e^- \to \ {\rm hadrons})$ ratio $R(s)$, 
we apply the mentioned evaluation to the effective charge 
${\cal F}(Q^{'2}) = d(Q^{'2})$ ($= \A(Q^{'2}) + {\cal O}(\A_2)$) of the Adler function
$d(Q^{'2})$, for complex $Q^{'2} = s \exp(i \phi)$, and integrate this 
expression in the contour integral (\ref{TF}).

Another example is the effective charge $r_{\tau}$ of the
strangeless $V+A$ semihadronic $\tau$ decay ratio $R_{\tau}$.
After removing the effects of nonzero quark masses, this quantity can be
expressed in terms of the effective charge of the Adler function
$d_{\rm Adl}(Q^2)$, defined in Eqs.~(\ref{ddef})-(\ref{PiV}), as the following contour integral:
\cite{Braaten:1988hc,Beneke:2008ad}:
\be
r_{\tau} = \frac{1}{2 \pi} \int_{-\pi}^{+ \pi}
d \phi \ (1 + e^{i \phi})^3 (1 - e^{i \phi}) \
d_{\rm Adl} (Q^2=m_{\tau}^2 e^{i \phi}) \ .
\label{rtaucont}
\ee

The mass-dependent timelike observables can be evaluated analogously as
the mentioned mass-independent ones. Furthermore, we can encounter
cases of such observables which have pQCD expansion in
noninteger powers [cf.~Eqs.~(\ref{disptAn3})-(\ref{AnutAnu})], 
such as is the
partial decay width of the Higgs to $b \bar b$; for an application
in this case, cf.~Ref.~\cite{GCAK} in a general IR fixed-point
framworks, and Ref.~\cite{BMS2} in the fractional APT (FAPT).

\section{Summary}
\label{summ}

We considered specific aspects of the problem of 
evaluation of QCD effects at hadronic (low) momenta 
$\alt 1$ GeV. Most of the lattice calculations and 
calculations using Dyson-Schwinger equations and/or other functional
methods indicate that the QCD running coupling freezes to
a finite value at low momenta, i.e., that it has
an IR fixed point. Various models with IR fixed point were 
considered here. We argued, on theoretical grounds,
that the perturbation expansions
of low-momentum spacelike physical quantities can be
used in such IR fixed point frameworks, provided that
the naive power series is abandoned and replaced by the
corresponding nonpower series in logarithmic derivatives.
We then numerically compared both types of expansion in
the case of the massless Adler function, for three different
frameworks of IR fixed point scenario. The numerical
results consistently indicate that the mentioned
nonpower expansion should be used, in order to have
nonperturbative effects at higher orders incorporated
correctly. The correct incorporation of these terms
results in a weaker renormalization scale dependence and 
in significantly better convergence properties of the series.

Moreover, a resummation method based on the (truncated) series in
logarithmic derivatives, namely a
generalization of the diagonal Pad\'e resummation method
proposed in Ref.~\cite{BGApQCD1} and later extended and applied
in IR fixed point scenarios \cite{GCRK,anOPE}, results in
very improved convergence properties of the 
evaluated series in all considered IR fixed point scenarios.

If, however, we work in the usual pQCD framework,
with the running coupling which suffers from the unphysical
(Landau) singularities, the mentioned methods of evaluations
for low-momentum spacelike physical quantities show no
improvement with respect to the usual truncated power series.

In addition, we argued how the described formalisms can be extended to
evaluations of low-energy timelike physical quantities.

On the other hand, it is unrealistic to assume that
the IR fixed point coupling is universal and, at the
same time, it gives us all the nonperturbative effects
via the mentioned evaluations. It appears to be more realistic to
assume that such a coupling, while being universal,
incorporates in the mentioned series only part of the
nonperturbative effects, and that other nonperturbative
effects can be added, for example via higher-twist terms
of OPE.\footnote{
On the other hand, approaches exist 
which eliminate the unphysical singularities, and
include nonperturbative effects, directly in
the specific (spacelike) obsevables, 
cf.~Refs.~\cite{DeRafael,MagrDual,Milton:2001mq,mes2,Nest3}
(see also \cite{Deur,Court}).}
If we want to apply OPE in such
IR fixed point scenarios, and if at the same time
we want to maintain the ITEP School interpretation of the OPE 
(that the higher-twist terms are exclusively of IR origin,
Refs.~\cite{Shifman:1978bx,DMW}), it is preferable that
the running coupling differs very little from the pQCD coupling at high 
squared momenta $Q^2$, by $\sim (\Lambda^2/Q^2)^N$ where $N$ is large.
Such models do exist, e.g., Refs.~\cite{1danQCD,2danQCD}
for $N=3, 5$, respectively; OPE in $\tau$ decay physics
has been applied with the model of Ref.~\cite{2danQCD}
in Ref.~\cite{anOPE}.

\begin{acknowledgments}
\noindent
This work was supported in part by FONDECYT (Chile) Grant No.~1130599.
\end{acknowledgments}

\appendix
\section{Subleading scheme dependence in scenarios with the coupling with
IR freezing}
\label{app1}

When we consider the subleading renormalization scheme dependence,
the basic pQCD relations are those of the
$c_j$-dependence of the pQCD coupling $a$, i.e., ``scheme RGEs''
\cite{Stevenson}
\bes
\label{RSchpRGEs}
\bea
\frac{\partial a}{\partial c_2} & = & a^3 + {\cal O}(a^5) 
\; \Rightarrow  \; \frac{\partial a^2}{\partial c_2} = 2 a^4 + \cdots \ ,
\\
\frac{\partial a}{\partial c_3} & = & \frac{1}{2} a^4 + \cdots \ .
\eea
\ees
For the IR fixed point scenarios (or any analytic model of $\A$), 
we can define the same
scheme dependence, under the correspondence $a^n \leftrightarrow \A_n$
\bes
\label{RSchanRGEs}
\bea
\frac{\partial \A}{\partial c_2} & = & \A_3 + {\cal O}(\A_5) \; 
\Rightarrow  \; \frac{\partial \A_2}{\partial c_2} = 2 \A_4 + \cdots \ ,
\\
\frac{\partial \A}{\partial c_3} & = & \frac{1}{2} \A_4 + \cdots \ .
\eea
\ees
These differential equations can be rewritten in terms of $\tA_n$'s using
the relations (\ref{As}). All the scheme dependence relations in
pQCD now carry over to IR fixed point scenarios, under the
correspondence  $a^n \leftrightarrow \A_n$ (or equivalently
$\ta_n \leftrightarrow \tA_n$)
\bes
\label{cjdep}
\bea
\frac{ \partial d_{\rm pt}^{[N]}}{\partial c_j}
&=& K^{(j)}_{N} a^{N+1}(\kappa Q^2) 
+  K^{(j)}_{N+1} a^{N+2}(\kappa Q^2) + \cdots \ , \ \Rightarrow
\label{cjdepDpt}
\\
\frac{ \partial d_{\rm an}^{[N]}}{\partial c_j}
&=& K^{(j)}_{N} \A_{N+1}(\kappa Q^2) 
+  K^{(j)}_{N+1} \A_{N+2}(\kappa Q^2) + \cdots \ ,
\label{cjdepDan}
\eea
\ees
and analogously for $d_{\rm man}^{[N]}$.

\section{Analogs of $a^n$ and ${\widetilde a}_n$ for noninteger $n=\nu$ in
IR fixed point scenarios}
\label{app2}

In the scenarios with IR freezing of the coupling $\A$,
for noninteger $n=\nu$ the quantities $\tA_{\nu}$ and $\A_{\nu}$
were obtained in Ref.~\cite{GCAK}. The idea was to perform
the analytic continuation of the most general form of the formulas for
$\tA_n$ and in $\A_n$ in $n$ ($\mapsto \nu$).

For this, a dispersion relation for the
logarithmic derivatives $\tA_{n+1}(Q^2)$ of Eq.~(\ref{tAn})
was obtained,
by applying the logarithmic derivatives
on the dispersion relation (\ref{A1disp}) for $\A(Q^2)$.
The obtained dispersion relation was then written in a form
involving the polylogarithm function of order $-n$
\be
\tA_{n+1}(Q^2) = \frac{1}{\pi} \frac{(-1)}{\beta_0^n \Gamma(n+1)}
\int_{0}^{\infty} \ \frac{d \sigma}{\sigma} \rho(\sigma)  
{\rm Li}_{-n} ( -\sigma/Q^2 ) \ ,
\label{disptAn2}
\ee
where we recall that $\rho(\sigma) \equiv {\rm Im} \A(-\sigma - i \epsilon)$.
Analytic continuation in $n \mapsto \nu$ then gives simply
\be
\tA_{\nu+1}(Q^2) = \frac{1}{\pi} \frac{(-1)}{\beta_0^{\nu} \Gamma(\nu+1)}
\int_{0}^{\infty} \ \frac{d \sigma}{\sigma} \rho(\sigma)  
{\rm Li}_{-\nu}\left( - \frac{\sigma}{Q^2} \right) \quad (-1 < \nu) \ ,
\label{disptAn3}
\ee
where $\nu$ can now be noninteger.
The couplings $\A_{\nu}$, which are (in IR fixed point scenario) analogs
of the noninteger powers $a^{\nu}$, 
can then be obtained as a linear combination
of the quantities $\tA_{\nu+m}$ ($m=0,1,2,\ldots$), via a generalization
of the relations (\ref{As}) to any integer $n$ and then replacing
$n \mapsto \nu$
\be
\A_{\nu} \equiv {\tA}_{\nu} + \sum_{m \geq 1}
\tk_m(\nu) {\tA}_{\nu + m} \quad (\nu > 0) \ .
\label{AnutAnu}
\ee
The coefficients $\tk_m(\nu)$ involve Gamma functions 
$\Gamma(x)$ and their derivatives (up to $m$ derivatives)
at the values $x=1, \nu+1, \nu+2, \ldots, \nu+m$, cf.~App.~A of Ref.~\cite{GCAK}.

It turns out that in the (fractional) APT model
of Refs.~\cite{ShS,MS,BMS1,BMS2,BMS3}, the (fractional) power analogs 
$\A_{\nu}^{\rm (APT)}$, Eq.~(\ref{MAAnudisp}),
constructed entirely from the discontinuities of the
pQCD coupling $a^{\nu}$, coincide with the result of
the approach described here, for the corresponding
special (APT) case: $\rho(\sigma) = \rho^{\rm (pt)}(\sigma)$, i.e., when
${\rm Im} \A(-\sigma - i \epsilon) = {\rm Im} a(-\sigma - i \epsilon)$.

\section{Distribution function of the leading-$\beta_0$ part of Adler function}
\label{appFd}

Here we review the formalism of the distribution function for the
leading-$\beta_0$ part of dimensionless renormalization scheme invariant
QCD quantities $d(Q^2)$, as developed in Ref.~\cite{Neubert}, 
but using here the
notations of the present paper. Further, an emphasis
will be made here which is slightly different from that of 
Ref.~\cite{Neubert}: namely, the distribution function $F_d(t)$, 
which generates the leading-$\beta_0$ parts of the perturbation 
coefficients of the quantity $d(Q^2)$, will appear in the
integral over momenta Eq.~(\ref{DLBint}) whose integrand includes 
the general ($N$-loop) coupling $a(t Q^2 e^{\cal C})$, 
and not necessarily one-loop coupling $a_{1\ell}(Q^2)$.
The formalism can be applied to spacelike and timelike
quantities. Throughout this Appendix, we can replace
the (usual) perturbative quantities $a$ and $\ta_{n+1}$ by
the (holomorphic) quantities of the IR fixed point scenarios 
$\A$ and $\tA_{n+1}$, cf.~Eqs.~(\ref{tan})-(\ref{tAn}).

In general, the coefficients $\td_n(\kappa)$ of the reorganized 
perturbation series of $d(Q^2)$, Eqs.~(\ref{Dmpt})-(\ref{Dman}), can be
written as a sum of powers of the number of active flavors $N_f$,
the latter indicating the number of loops of (massless) quark flavors
\bes
\label{tdnexp}
\bea
\td_n(\kappa) &=& {\cal C}_{n,n}(\kappa) N_f^n + 
{\cal C}_{n,n-1}(\kappa) N_f^{n-1}  + \ldots + {\cal C}_{n,0} \ ,
\label{tdnexpNf}
\\
&=& c_{n,n}(\kappa) \beta_0^n + c_{n,n-1} \beta_0^{n-1} + \ldots
+ c_{n,0} \ ,
\label{tdnexpb0}
\eea
\ees
where the leading-$N_f$ and the leading-$\beta_0$ (LB) coefficients
are simply related \cite{LTM}
\be
c_{n,n} = (-6)^n {\cal C}_{n,n} \ ,
\label{ccalC}
\ee
due to the relation $N_f = - 6 \beta_0 + 33/2$. It is interesting
that the dependence
on the renormalization scale $\mu$ ($\kappa \equiv \mu^2/Q^2$)
of the LB coefficient 
$\td_n^{\rm (LB)}(\kappa) = c_{n,n}(\kappa) \beta_0^n$ ($c_{0,0}=1$)
is the same  as for the entire coefficients $\td_n(\kappa)$ 
[cf.~Eqs.~(\ref{tdndiff})-(\ref{tdnmu})]
\bes
\label{cnnmu}
\bea
\frac{d c_{n,n}(\kappa)}{d \ln \kappa} &=& n c_{n-1,n-1}(\kappa)
\; \Rightarrow
\label{tdndiff2}
\\
c_{n,n}(\kappa_2) &=& \sum_{p=0}^n 
\left(
\begin{array}{c}
n \\
p
\end{array}
\right)
\ \ln^{n-p} \left( \frac{\kappa_2}{\kappa_1} \right) c_{p,p}(\kappa_1) \ ,
\label{tdnmu2}
\eea
\ees
(for the above equality, see, for example, Ref.~\cite{CV2}).
The LB part of the reorganized perturbation
series [cf.~Eqs.~(\ref{Dmpt})-(\ref{Dman})] is
\bea
d^{\rm (LB)}(Q^2)_{\rm mpt} &=& a(\kappa Q^2) + 
\sum_{n=1}^{\infty} c_{n,n}(\kappa) \; \beta_0^n \; {\ta}_{n+1}(\kappa Q^2) \ .
\label{DmptLB}
\eea
This expression can be written as an integral over the squared
momenta $-k^2=t Q^2$ of the form (\ref{DLBint})
\bea
d^{\rm (LB)}(Q^2)_{\rm mpt} & = &   
\int_0^{\infty} \frac{dt}{t} \; F_d(t) \; a(t Q^2 e^{{\cal C}}) \ ,
\label{DLBint2}
\eea
where ${\cal C}=-5/3$ in $\MSbar$ scheme (scaling convention).
We now review how the distribution function $F_d(t)$ which
appears in this integral is obtained
from the knowledge of the LB coefficients $c_{n,n}$, following
Ref.~\cite{Neubert} (with adapted notations).

If in the LB-series (\ref{DmptLB}) the logarithmic derivatives
${\ta}_{n+1}(\kappa Q^2)$ are formally replaced by powers
$a(\kappa Q^2)^{n+1}$ (note: ${\ta}_{n+1} \not= a^{n+1}$ if general $N$-loop
running with $N \geq 2$), we name the new formal LB quantity as $\td^{\rm (LB)}$
\bea
\td^{\rm (LB)}(Q^2;\kappa^2)^{\rm (LB)} & \equiv & a(\kappa Q^2) + 
\sum_{n=1}^{\infty} c_{n,n}(\kappa) \; \beta_0^n a^{n+1}(\kappa Q^2) \ .
\label{tDmptLB}
\eea
It is in general renormalization scale dependent, unless $a(\mu^2)$ runs
according to one-loop. The Borel transform of this quantity is
\bes
\label{Btd}
\bea
B_{\td}(b;\kappa)&=& 1 + \frac{\td_1^{\rm (LB)}(\kappa)}{1!\beta_0} b +
\frac{\td_2^{\rm (LB)}(\kappa)}{2!\beta_0^2} b^2 + \ldots 
\label{Btd1}
  \\
&=& 1 + \frac{c_{1,1}(\kappa)}{1!} b +  \frac{c_{2,2}(\kappa)}{2!} b^2 + \ldots \ .
\label{Btd2}
\eea
\ees
The idea is to relate this quantity with the distribution function
$F_d(t)$ via an integral relation, and to invert that relation in
order to obtain $F_d$.
The scale dependence of the quantity (\ref{Btd}) 
is obtained immediately from the relations (\ref{tdndiff2})
\be
B_{\td}(b;\kappa) =  B_{\td}(b;1) \; \kappa^b \ .
\label{BtdRScl}
\ee
However, the renormalization scale $\mu^2$ ($\equiv \kappa Q^2$) has
scaling convention dependence ($\Lambda$ definition), 
also called $\Lambda$ scheme or ${\cal C}$-dependence.
This is reflected by the fact that the RGE running of
the coupling $a(\mu^2)$ is scaling convention dependent
\cite{Stevenson}, while the quantity
$a(\mu^2 e^{\cal C})$ has no such dependence, where 
${\cal C}=-5/3$, $-5/3 + \gamma_{\rm E} - \ln(4 \pi)$, and $0$, in the
scaling convention frameworks $\MSbar$, MS, and V, respectively.
Specifically, we have $a(\mu^2) = f(\mu^2/\Lambda^2)$ where $\Lambda$
is different in different scaling conventions, and we have the
following relations:
%\end{document}
\bes
\label{Cdep}
\bea
\mu^2 &=& \mu_{(0)}^2 e^{\cal C} \ , \quad 
\Lambda^2 = \Lambda_{(0)}^2 e^{\cal C} \;  \Rightarrow
\label{Cdep1}
\\
a(\kappa Q^2) &=& a \left( \kappa e^{-{\cal C}} Q^2; {\cal C}=0 \right) \ ,
\quad
\ta_{n+1}(\kappa Q^2) = 
\ta_{n+1}\left( \kappa e^{-{\cal C}} Q^2; {\cal C}=0 \right) \ ,
\label{Cdep2}
\\
\td_n(\kappa) & = & \td_n \left( \kappa e^{-{\cal C}}; {\cal C}=0 \right) \ ,
\qquad
c_{n,n}(\kappa) = c_{n,n}\left( \kappa e^{-{\cal C}}; {\cal C}=0 \right) \ ,
\label{Cdep3}
\eea
\ees
where, as usual, $\kappa \equiv \mu^2/Q^2$, and the subscript ``$(0)$'' denotes
the V scaling convention (${\cal C}=0$). The last identity implies that
the Borel transform (\ref{Btd2}) has, in addition to the $\kappa$-dependence
(\ref{BtdRScl}), also ${\cal C}$-dependence
\be
B_{\td}(b;\kappa) =B_{\td} \left( b;\kappa e^{-{\cal C}}; {\cal C}=0 \right)
= B_{\td}(b; 1; {\cal C}=0) (\kappa e^{-{\cal C}})^b =
 B_{\td}(b; e^{\cal C})  (\kappa e^{-{\cal C}})^b \ .
\label{BtdRSclCd}
\ee
This leads to the following definition of the renormalization scale
independent and ${\cal C}$-independent Borel transform
${\hat B}_{\td}(b)$:
\bes
\label{hatB}
\bea
{\hat B}_{\td}(b) &\equiv& B_{\td}(b; \kappa) 
(\kappa e^{-{\cal C}})^{-b} 
\label{hatB1}
\\
&=& B_{\td}(b; \kappa=e^{\cal C}) 
 = 1 + \frac{c_{1,1}(e^{\cal C})}{1!} b +  
\frac{c_{2,2}(e^{\cal C})}{2!} b^2 + \ldots
\ .
\label{hatB2}
\eea
\ees

In the integral (\ref{DLBint2}) the
(${\cal C}$-independent) quantity $a(t Q^2 e^{{\cal C}})$ 
can be expanded\footnote{
If $a(t Q^2 e^{{\cal C}})$ RGE-evolves at a $N$-loop level with $N > 2$, 
then $a(t Q^2 e^{{\cal C}})$ and thus the quantity $d^{\rm (LB)}(Q^2)$
acquire the renormalization scheme dependence, i.e.,
dependence on the beta-parameters $c_j \equiv \beta_j/\beta_0$
($j=2,\ldots,N-1$).}
around the point $\ln \mu^2$, using the definitions (\ref{tan})
\be
a(t Q^2 e^{{\cal C}}) = a(\mu^2) + (-\beta_0) \ln \left(
t \frac{Q^2}{\mu^2} e^{\cal C} \right) \ta_2(\mu^2)
+  (-\beta_0)^2 \ln^2 \left(
t \frac{Q^2}{\mu^2} e^{\cal C} \right) \ta_3(\mu^2) + \ldots \ .
\label{aexp}
\ee
Using this in Eq.~(\ref{DLBint2}) and exchanging there the order of 
summation and integration,
leads to the series Eq.~(\ref{DmptLB}), with the
following relation between the LB-coefficientes $c_{n,n}$
and the searched for distribution function $F_d$ 
($c_{n,n}$ are assumed known)
\be
c_{n,n}(\kappa) = (-1)^n \int_{t=0}^{\infty} d (\ln t)
\ln^n \left( t \kappa^{-1} e^{\cal C} \right) F_d(t) \ ,
\label{cnns}
\ee
where $n=0,1,2, \ldots$ and $c_{0,0}=1$ 
[when $\kappa=1$ this gives Eq.~(\ref{tdns})].
Inserting these expressions into the expansion 
(\ref{Btd}) of the Borel transform $B_{\td}(b;\kappa)$,
and exchanging the order of integration and summation,
leads to 
\bea
B_{\td}(b;\kappa) & = & \int_{t=0}^{\infty} d (\ln t) \;
F_d(t) \left[ 1 + 
\sum_{n=1}^{\infty} \frac{1}{n!} (-1)^n \ln^n(t \kappa^{-1} e^{\cal C}) b^n \right] = (\kappa e^{-{\cal C}})^b \int_0^{\infty} 
dt F_d(t) t^{-b-1} \ ,
\label{Mell}
\eea
which can be rewritten in terms of the invariant Borel transform
(\ref{hatB}) as
\be
{\hat B}_{\td}(t)=  \int_0^{\infty}  dt F_d(t) t^{-b-1} \ .
\label{Mellhat}
\ee
This is the Mellin transform of $F_d$, 
and the inverse transformation then gives
the distribution function $F_d$
\be
F_d(t) = \frac{1}{2 \pi i} \int_{c - i \infty}^{c + i \infty}
db \; {\hat B}_{\td}(t) t^b \ ,
\label{Mellhatinv}
\ee
where the real value $c$ is any value between the first UV and the
first IR renormalon pole of $d(Q^2)$ in $b$ plane: $-k < c < K$ (for
the Adler function: $-k=-1$, $K=2$). Therefore, if the
LB coefficients $c_{n,n}$ and thus ${\hat B}_{\td}(t)$ are known,
Eq.~(\ref{Mellhatinv}) gives the distribution function $F_d$ of the
quantity $d^{\rm (LB)}(Q^2)$ [Eq.~(\ref{DmptLB})] which appears
in the momentum-integral (\ref{DLBint2}) and is the complete
LB part of the full quantity $d(Q^2)$.

For the QCD Adler function, the large-$N_f$ 
coefficients ${\cal C}_{n,n}$, Eq.~(\ref{tdnexpNf}), with
$\kappa=e^{\cal C}$,
can be deduced from Ref.~\cite{Broad1} where their generating function
was obtained in the context of large-$N$ expansion in QED.\footnote{
Cf.~also Ref.~\cite{Broad2} where ${\cal C}_{n,n}$ were obtained
for the QCD Adler function explicitly in $\MSbar$ scheme and at $\kappa=1$. 
The LB coefficents $c_{n,n}(\kappa=1)$ are then obtained by Eq.~(\ref{ccalC}),
and then the coefficients $c_{n,n}(\kappa=e^{\cal C})$ (${\cal C}=-5/3$)
by (\ref{cnnmu}).}
The LB coefficents $c_{n,n}$ are then obtained by Eq.~(\ref{ccalC}), giving 
\be
c_{n,n}(e^{\cal C}) = \frac{3}{4} C_F \left( \frac{d}{db} \right)^n
P(1-b)|_{b=0} \ ,
\label{cnnAdl}
\ee
where $C_F=(N_c^2-1)/(2 N_c) = 4/3$, and $P(x)$ is the trigamma function 
obtained in Ref.~\cite{Broad1}\footnote{It can be expressed as
a combination of the Hurwitz zeta functions $\xi(s,y)$ with $s=2$:
\begin{displaymath}
P(x)= \frac{2}{3} \frac{1}{x(1+x)}\left[ \xi\left( 2,\frac{1}{2}(2-x) \right)-\xi \left( 2,\frac{1}{2}(3-x) \right) 
- \xi \left( 2, \frac{1}{2}(2+x) \right) + 
\xi\left( 2,\frac{1}{2}(3+x) \right) \right] \ .
\end{displaymath}}
by extension of the analysis of Ref.~\cite{Palanq}
\be
P(x) = \frac{32}{3 (1 + x)} \sum_{k=2}^{\infty}
\frac{(-1)^k k}{(k^2-x^2)^2} \ .
\label{trigamma}
\ee
Eq.~(\ref{cnnAdl}) gives, via Eq.~(\ref{hatB2}), 
the invariant Borel transform for the LB Adler function
\be
{\hat B}_{\td}(b)  = \frac{3}{4} C_F P(1 - b) \ .
\label{hatBAdl}
\ee
The function $P(x)$ is thus the generator of the 
Adler function LB coefficients $c_{n,n}(\kappa)$ with $\kappa$ ($ \equiv \mu^2/Q^2$)
$= e^{\cal C}$
\bea
P(1 - b) &=&  1 + \left( \frac{23}{6} - 4 \xi_3 \right) b +
\frac{1}{2!} 6 \left( 3 - 2 \xi_3 \right)  b^2 + \frac{1}{3!}
\left( \frac{201}{2} - 42 \xi_3 - 60 \xi_5 \right) b^3 
\nonumber\\
&& + \left( \frac{1305}{2} - 180 \xi_3 - 360 \xi_5 \right) b^4 
+ \ldots
\ .
\label{Pexp}
\eea
Using the results (\ref{trigamma})-(\ref{hatBAdl}) in Eq.~(\ref{Mellhatinv})
with $c=1$, and $z=-i(1-b)$, and exchanging the order of integration and summation,
leads to 
\bea
F_d(t) &=& \frac{4 C_F}{\pi} t \; \sum_{k=2}^{\infty} (-1)^k \int_{-\infty}^{\infty}
dz \; \exp(- i z \ln t) \frac{1}{1+i z} \frac{k}{(k^2 + z^2)^2}
\nonumber\\
&=& i \frac{2 C_F}{\pi} t \; \sum_{k=2}^{\infty} (-1)^k \left( \frac{d}{d k} \right)
\int_{-\infty}^{\infty} dz \; \exp(- i z \ln t) \frac{1}{(z-i) (z-i k) (z+ik)} \ .
\label{FdAdl1}
\eea
In the integration over $z$ the Cauchy formula can be used: when $0<t<1$, 
the integration path is closed in the upper half plane; when $t>0$, it is closed
in the lower half plane. This gives the following result for the (LB) Adler
distribution function:
\bes
\label{FdAdlres}
\bea
F_d(t)_{(t<1)} &=& 2 C_F t \sum_{k=2}^{\infty}  \left[ 
t (-1)^k \left( \frac{1}{(k-1)^2} - \frac{1}{(k+1)^2} \right) 
+\ln (t) (-1)^k t^k \left( \frac{1}{(k-1)} - \frac{1}{k} \right)
- (-1)^k t^k \left( \frac{1}{(k-1)^2} - \frac{1}{k^2} \right) \right]
\nonumber\\
&=& 2 C_F t\left[-t \ln (t) + (1+t) \ln(1+t) \ln (t) + \frac{7}{4} t + (1+t)
{\rm Li}_2(-t) \right] \ ,
\label{FdAdltsm1}
\\
F_d(t)_{(t>1)} &=& 2 C_F t \sum_{k=2}^{\infty} \left[ 
\ln (t) \frac{(-1)^k}{t^k} \left( \frac{1}{k} - \frac{1}{(k+1)} \right) 
+ \frac{(-1)^k}{t^k} \left( \frac{1}{k^2} - \frac{1}{(k+1)^2} \right )
\right]
\nonumber\\
&=& 2 C_F \left[\left( \frac{1}{2} + t \right) \ln (t) - t (1+t)  \ln (t) \left( 1+ \frac{1}{t} \right) 
+ \left( \frac{3}{4} +  t \right)  +  t (1+t)
{\rm Li}_2 \left( -\frac{1}{t} \right) \right] \ .
\label{FdAdltsm2} 
\eea
\ees
Use has been made of the expansion of the polylogarithm function
\be
{\rm Li}_2(-t) = \sum_{k=1}^{\infty} (-1)^k \frac{t^k}{k^2} \qquad (|t|<1) \ .
\label{poly}
\ee
As mentioned, the result (\ref{FdAdlres}) was first obtained in Ref.~\cite{Neubert},
the funtion ${\hat w}(\tau)$ there is $4 F_d(\tau)/\tau$ here.
The integration (\ref{DLBint2}) is applied here, however, for a general
$N$-loop coupling $a(t Q^2 e^{\cal C})$; in the case of IR fixed point scenarios,
$a \mapsto \A$ and the integral is now convergent, apart from being
renormalization scale and ${\cal C}$-independent. 
If the usual $\MSbar$-like coupling
 $a_{\rm pt}(t Q^2 e^{\cal C})$  is applied, the integral becomes ambiguous due
to the Landau (cut and pole) singularities of  $a_{\rm pt}(t Q^2 e^{\cal C})$ at
low $t$, and an integration prescription is necessary; in such cases, we use the
principal value prescription
\bea
d^{\rm (LB)}(Q^2)_{\rm (m)pt} & = &   
{\rm Re} \int_{+i \epsilon}^{+i \epsilon + \infty} 
\frac{dt}{t} \; F_d(t) \; a(_{\rm pt}t Q^2 e^{{\cal C}}) \ ,
\quad (\epsilon \to +0) \ .
\label{DLBint3}
\eea 

\section{Exact solutions of RGE in terms of Lambert function}
\label{appLamb}

In the scheme where $c_2=c_3=\ldots = 0$,
the pQCD running coupling $a(\kappa Q^2)$
has formally the two-loop form and there is an exact solution of
the RGE in this case, cf.~Refs.~\cite{Gardi:1998qr,Magr}\footnote{
The authors of Ref.~\cite{GarKat} demonstrated that in this scheme the
beta function does not factor out in the generalized Crewther relation.}
\bea
a(\kappa Q^2) = - \frac{1}{c_1} \frac{1}{\left[
1 + W_{\mp 1}(z) \right]} \ .
\label{aptexact}
\eea
Here, $Q^2=|Q^2| \exp(i \phi)$; $W_{-1}$ and $W_{+1}$
are the branches of the Lambert function
for $0 \leq \phi < + \pi$ and $- \pi < \phi < 0$, 
respectively, and $z$ is defined as
\be
z =  - \frac{1}{c_1 e} 
\left( \frac{\kappa |Q^2|}{\Lambda_{\rm L.}^2} \right)^{-\beta_0/c_1} 
\exp \left( - i {\beta_0}\phi/c_1 \right) \ ,
\label{zexpr}
\ee 
where $\Lambda_{\rm L.}$ is the Lambert QCD scale.\footnote{ 
$\Lambda_{\rm L.} \sim {\Lambda}$, where ${\Lambda}$ is the usual $\MSbar$ scale
appearing in the expansion of pQCD coupling $a(Q^2)$ in inverse powers
of logarithms ${\cal L} \equiv \ln(Q^2/\Lambda^2)$: $a(Q^2) = 1/(\beta_0 {\cal L}) - (c_1/\beta_0^2)
(\ln {\cal L}/{\cal L}^2) + {\cal O}(\ln^2 {\cal L}/{\cal L}^3)$. It can be checked that in the $c_2=c_3=\cdots=0$ renormalization scheme with $N_f=3$ we have: $\Lambda_{\rm L.} = \Lambda/0.72882$. We use $\Lambda_{\rm L.}=0.487$ GeV, thus $\Lambda=0.355$ GeV.}

In the analytic QCD model with two deltas (2$\delta$anQCD), 
Ref.~\cite{2danQCD}, however, 
we use for the renormalization scheme the 
preferred central Lambert scheme of the model (with $N_f=3$):
$c_2 = -4.76$, $c_j = c_2^{j-1}/c_1^{j-2}$ ($j=3,4,\ldots$).
The exact solution of the underlying pQCD coupling is also known in
this case, again in terms of the Lambert function 
(Ref.~\cite{Gardi:1998qr}, cf.~also
Ref.~\cite{CveKon}). 
\bea
a(\kappa Q^2) = - \frac{1}{c_1} \frac{1}{\left[
1 - c_2/c_1^2 + W_{\mp 1}(z) \right]} \ , \quad 
\left( \Lambda_{\rm L.}=0.260 \ {\rm GeV} \ ; \; 
\rho^{\rm (pt)}(\sigma) = {\rm Im} \; a(-\sigma - i \epsilon) \right) \ ,
\label{aptLambexact}
\eea
and $z$ is again defined by Eq.~(\ref{zexpr}).

\end{document}